\documentclass[pre,showpacs,preprintnumbers,floatfix,amsmath,amssymb,preprint]{revtex4}

\usepackage{graphicx}
\usepackage{latexsym}
\usepackage{amsmath}
\usepackage{amssymb}
\usepackage{amsfonts}
\usepackage{bm}

\advance\voffset 1.5cm

\newcommand{\la}{\left<}
\newcommand{\ra}{\right>}
\newcommand{\ddiff}{\ensuremath{\text{d}}}

\newcommand{\qvec}{\ensuremath{\underline{q}}}
\newcommand{\rvec}{\ensuremath{\underline{r}}}

\newcommand{\Trace}{\ensuremath{\text{Tr}}}

\newcommand{\xijl}{\ensuremath{x_{l}}}
\newcommand{\yijl}{\ensuremath{y_{l}}}
\newcommand{\rijl}{\ensuremath{r_{l}}}
\newcommand{\rijlvec}{\ensuremath{\underline{r}_{l}}}
\newcommand{\fijl}{\ensuremath{f_{l}}}
\newcommand{\rijlprime}{\ensuremath{r_{l^{\prime}}}}

\newcommand{\kB}{\mbox{$k_{\rm B}$}}
\newcommand{\kBT}{\mbox{$k_{\rm B}T$}}
\newcommand{\nmon}{\ensuremath{n_\mathrm{mon}}}
\newcommand{\etotal}{\ensuremath{e_\mathrm{tot}}}
\newcommand{\ebond}{\ensuremath{e_\mathrm{b}}}
\newcommand{\enb}{\ensuremath{e_\mathrm{nb}}}
\newcommand{\einter}{\ensuremath{e_\mathrm{int}}}
\newcommand{\Pbond}{\ensuremath{P_\mathrm{b}}}
\newcommand{\Pnb}{\ensuremath{P_\mathrm{nb}}}
\newcommand{\Pinter}{\ensuremath{P_\mathrm{int}}}
\newcommand{\Pex}{\ensuremath{P_\mathrm{ex}}}
\newcommand{\Pid}{\ensuremath{P_\mathrm{id}}}

\newcommand{\aP}{\ensuremath{a_\mathrm{P}}}
\newcommand{\Wvirial}{\ensuremath{{\cal W}}}
\newcommand{\kappaT}{\ensuremath{\kappa_\mathrm{T}}}
\newcommand{\gT}{\ensuremath{g_\mathrm{T}}}

\newcommand{\gTN}{\ensuremath{g_\mathrm{T,N}}}
\newcommand{\Xvirial}{\ensuremath{\eta_\mathrm{B}}}
\newcommand{\Xvirialbond}{\ensuremath{\eta_\mathrm{B,b}}}
\newcommand{\Xvirialnb}{\ensuremath{\eta_\mathrm{B,nb}}}
\newcommand{\Xfluctu}{\ensuremath{\eta_\mathrm{F}}}
\newcommand{\Xfluctubond}{\ensuremath{\eta_\mathrm{F,b}}}
\newcommand{\Xfluctunb}{\ensuremath{\eta_\mathrm{F,nb}}}
\newcommand{\Xfluctuself}{\ensuremath{\eta_\mathrm{F,self}}}
\newcommand{\Xfluctudist}{\ensuremath{\eta_\mathrm{F,dist}}}
\newcommand{\Xfluctuselfbond}{\ensuremath{\eta_\mathrm{F,self,b}}}
\newcommand{\Xfluctuselfnb}{\ensuremath{\eta_\mathrm{F,self,nb}}}
\newcommand{\Xfluctumix}{\ensuremath{\eta_\mathrm{F,mix}}}
\newcommand{\muB}{\ensuremath{{\mu_\mathrm{B}}}}
\newcommand{\laB}{\ensuremath{{\lambda_\mathrm{B}}}}
\newcommand{\muF}{\ensuremath{{\mu_\mathrm{F}}}}
\newcommand{\laF}{\ensuremath{{\lambda_\mathrm{F}}}}

\newcommand{\Tcrit}{\ensuremath{T_\mathrm{c}}}

\newcommand{\Lbox}{\ensuremath{L_\mathrm{box}}}
\newcommand{\ubond}{\ensuremath{u_{\text{b}}}}
\newcommand{\lbond}{\ensuremath{l_{\text{b}}}}
\newcommand{\kbond}{\ensuremath{k_{\text{b}}}}
\newcommand{\Tbond}{\ensuremath{\tau_{\text{b}}}}
\newcommand{\kstiff}{\ensuremath{k_{\theta}}}

\newcommand{\dperi}{\ensuremath{d_\mathrm{p}}}
\newcommand{\thetatwo}{\ensuremath{\theta_\mathrm{2}}}
\newcommand{\nudil}{\ensuremath{\nu_\mathrm{0}}}
\newcommand{\thetadil}{\ensuremath{\theta_\mathrm{2,0}}}

\newcommand{\bdil}{\ensuremath{b_\mathrm{0}}}
\newcommand{\benddil}{\ensuremath{b_\mathrm{e,0}}}
\newcommand{\bgyrdil}{\ensuremath{b_\mathrm{g,0}}}
\newcommand{\Rend}{\ensuremath{R_\text{e}}}
\newcommand{\Rgyr}{\ensuremath{R_\mathrm{g}}}

\newcommand{\Rdil}{\ensuremath{R_\mathrm{0}}}
\newcommand{\rhostar}{\ensuremath{\rho^*}}
\newcommand{\rhostarstar}{\ensuremath{\rho^{**}}}
\newcommand{\dcm}{\ensuremath{d_\text{cm}}}

\newcommand{\ninter}{\ensuremath{n_\mathrm{int}}}

\newcommand{\Fdil}{\ensuremath{F_\mathrm{0}}}
\newcommand{\cdil}{\ensuremath{c_\mathrm{0}}}

\begin{document}
\title{Strictly two-dimensional self-avoiding walks:\\ Thermodynamic properties revisited}

\author{N.~Schulmann}
\affiliation{Institut Charles Sadron, Universit\'e de Strasbourg \& CNRS, 23 rue du Loess, BP 84047, 67034 Strasbourg Cedex 2, France}
\author{H. Xu}
\affiliation{LCP-A2MC, Institut Jean Barriol, Universit\'e de Lorraine \& CNRS, 1 bd Arago, 57078 Metz Cedex 03, France}
\author{H. Meyer}
\affiliation{Institut Charles Sadron, Universit\'e de Strasbourg \& CNRS, 23 rue du Loess, BP 84047, 67034 Strasbourg Cedex 2, France}
\author{P.~Poli\'nska}
\affiliation{Institut Charles Sadron, Universit\'e de Strasbourg \& CNRS, 23 rue du Loess, BP 84047, 67034 Strasbourg Cedex 2, France}
\author{J. Baschnagel}
\affiliation{Institut Charles Sadron, Universit\'e de Strasbourg \& CNRS, 23 rue du Loess, BP 84047, 67034 Strasbourg Cedex 2, France}
\author{J.P.~Wittmer}
\email{joachim.wittmer@ics-cnrs.unistra.fr}
\affiliation{Institut Charles Sadron, Universit\'e de Strasbourg \& CNRS, 23 rue du Loess, BP 84047, 67034 Strasbourg Cedex 2, France}

\begin{abstract}
The density crossover scaling of various thermodynamic properties of solutions and melts
of self-avoiding and highly flexible polymer chains without chain intersections confined to 
strictly two dimensions is investigated by means of molecular dynamics and Monte Carlo 
simulations of a standard coarse-grained bead-spring model. 
In the semidilute regime we confirm over an order of magnitude of the monomer density $\rho$ 
the expected power-law scaling for the interaction energy between different chains $\einter \sim \rho^{21/8}$, 
the total pressure $P \sim \rho^3$ and
the dimensionless compressibility $\gT = \lim_{q \to 0} S(q) \sim 1/\rho^2$.
%
Various elastic contributions associated to the affine and non-affine response to
an infinitesimal strain are analyzed as functions of density and sampling time.
We show how the size $\xi(\rho)$ of the semidilute blob may be determined experimentally 
from the total monomer structure factor $S(q)$ characterizing the compressibility of the
solution at a given wavevector $q$.
%
We comment briefly on finite persistence length effects.
\end{abstract}


\date{\today}

\maketitle

\section{Introduction}
\label{sec_intro}

\begin{figure}[t]
\centerline{\resizebox{0.8\columnwidth}{!}{\includegraphics*{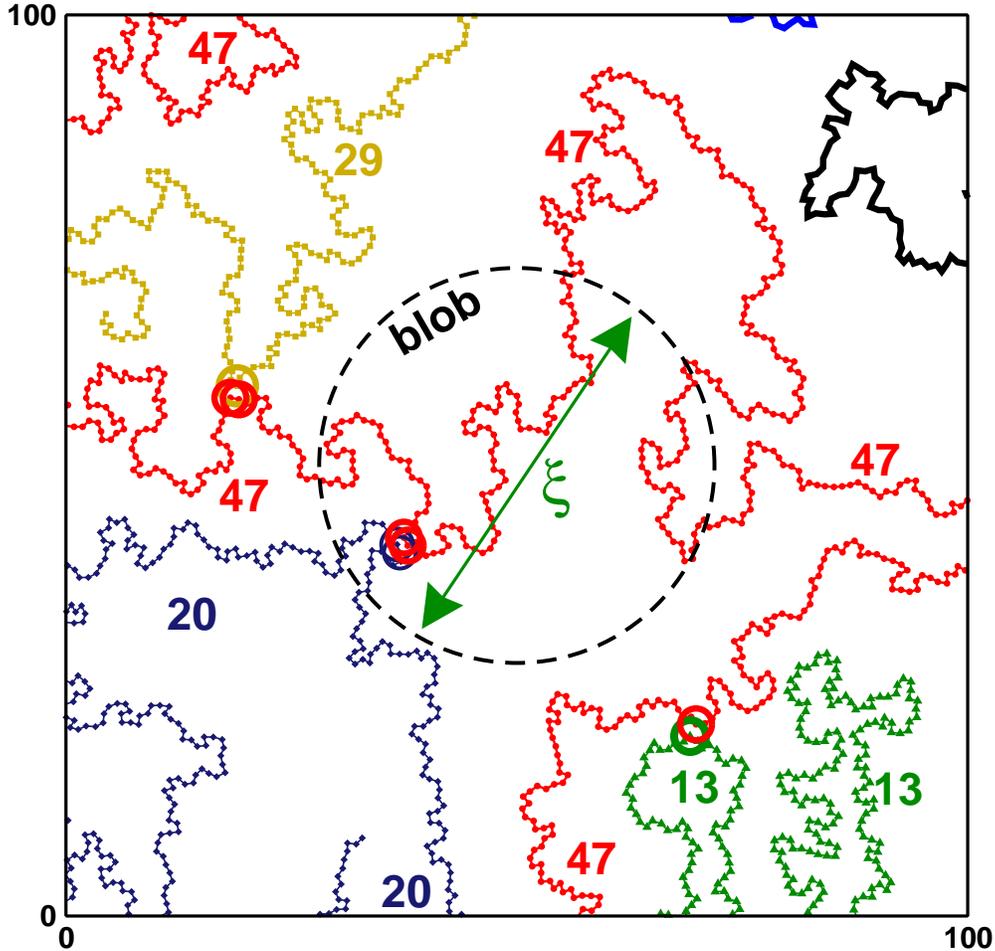}}}
\caption{We consider numerically thermodynamic properties of self-avoiding polymers in strictly 
two dimensions without chain intersections focusing on the experimentally relevant semidilute regime. 
The snapshot has been obtained for our coarse-grained model at number density $\rho=0.125$ and chain length $N=2048$. 
The numbers refer to a chain index used for computational purposes. 
The large dashed circle represents a hard disk of uniform mass distribution having a radius of gyration 
$\Rgyr$ equal to the semidilute blob size $\xi \approx 31$ of the given density. 
While the chains adopt compact configurations on larger distances $r \gg \xi$ 
as characterized by the exponents $\nu =  1/d$ and $\thetatwo = 3/4$, 
the swollen chain statistics ($\nudil = 3/4$, $\thetadil=19/12$)
remains relevant on smaller scales as shown in ref.~\cite{SMW12}.
The few small open circles (not to scale) correspond to monomers interacting with monomers from other chains.
As shown in sect.~\ref{res_energy}, the interaction energy $\einter$ due to such contacts remains small, 
even at higher densities, if the chain length is large. 
\label{fig_snap_semi}
}
\end{figure}

\paragraph*{Compact chains of fractal perimeter.}
Dense polymer solutions confined to effectively two-dimensional (2D) thin layers are
of significant technological relevance with opportunities ranging from tribology to biology
\cite{RLJones1999,Granick03,McKenna2005}. We focus here on the conceptionally important limit 
of self-avoiding homopolymers confined to strictly $d=2$ dimensions where chain intersections 
are forbidden as shown by the snapshot presented in fig.~\ref{fig_snap_semi}. Such systems
are not only of theoretical \cite{DegennesBook,Duplantier89,ANS03} and computational
\cite{carmesin90,Rutledge97,Yethiraj03,Cavallo03,Cavallo05,MKA09,MWK10,MSZ11,SMW12,WMJ10}
but also of experimental interest \cite{mai00,Deutsch05,Rubinstein07,Monroy05,Monroy07,Monroy10,Monroy11,Kumaki12},
especially since conformational properties may directly be visualized \cite{mai00,Rubinstein07,Kumaki12}. 
It is now generally accepted
\cite{DegennesBook,Duplantier89,ANS03,mai00,carmesin90,Rutledge97,Yethiraj03,Cavallo03,Cavallo05,MKA09,MWK10,MSZ11,SMW12}
that at sufficiently high monomer density $\rho$ and chain length $N$
the chains adopt compact conformations,
i.e., the typical chain size $R$ scales as \cite{foot_approx}
\begin{equation}
R \approx (N/\rho)^{\nu} \mbox{ where } \nu=1/d=1/2.
\label{eq_compact}
\end{equation}
We stress that eq.~(\ref{eq_compact}) does {\em not} imply Gaussian chain statistics 
since other critical exponents with non-Gaussian values have
been shown to matter for various experimentally relevant properties \cite{Duplantier89,ANS03,MWK10}.
It is thus incorrect to assume that excluded-volume effects are screened \cite{Cavallo03}
as is approximately the case for three-dimensional melts \cite{WCX11}.
Interestingly, compactness does not imply disklike shapes minimizing the chain perimeter length $L$.
In fact the perimeter is found to be fractal with \cite{ANS03,MKA09,MWK10,MSZ11,SMW12}
\begin{equation}
L \sim N \ninter \sim R^{\dperi} \sim N^{\dperi/d}
\mbox{ with } \dperi = d - \thetatwo = 5/4
\label{eq_dperi}
\end{equation}
where $\ninter$ stands for the fraction of monomers interacting with other chains and 
the fractal line dimension $\dperi$ is set by Duplantier's contact exponent
$\thetatwo=3/4$ \cite{Duplantier89}.

\paragraph*{Semidilute regime.}
Duplantier's predictions obtained using conformal invariance \cite{Duplantier89}
rely on the non-intersection constraint and the space-filling property of the melt.
Obviously, high densities are experimentally difficult to realize for strictly 2D layers
\cite{mai00,Deutsch05} since chains tend either to detach or to overlap,
increasing thus the number of layers as demonstrated from the pressure
isotherms studied in ref.~\cite{Deutsch05}. It is thus of some importance that eqs.~(\ref{eq_compact},\ref{eq_dperi})
have been argued to hold more generally for all densities assuming that the chains are sufficiently long \cite{SMW12}.
Following de Gennes' classical density crossover scaling \cite{DegennesBook}
polymer solutions may be viewed as space-filling melts of ``blobs" of size $\xi(\rho)$
containing $g(\rho) \approx \rho \xi^d(\rho)$ monomers with
\begin{eqnarray}
\xi(\rho) & \approx & \bdil g^{\nudil} \sim 1/\rho^{3/2} \mbox{ and } \nonumber \\
g(\rho) & \approx & 1 /(\bdil^d \rho)^{1/(\nudil d -1)} \sim 1/\rho^2.
\label{eq_grho}
\end{eqnarray}
Here $\nudil=3/4$ stands for Flory's chain size exponent for dilute swollen chains in $d=2$ dimensions
\cite{DegennesBook} and $\bdil \equiv \lim_{N\to \infty} \Rdil(N)/N^{\nudil}$ for the corresponding
statistical segment size. 
(Throughout the paper the dilute limit of a property is often characterized by an index $0$.)
Focusing on conformational properties it has thus been confirmed numerically 
that, e.g., $\ninter \sim 1/N^{\nu \thetatwo}=1/N^{3/8}$ for all densities if $N \gg g(\rho)$
while $\ninter \sim 1/N^{\nudil \thetadil} = 1/N^{19/16}$ for $N \ll g(\rho)$
with $\thetadil = 19/12$ being the contact exponent in the dilute limit.
We remind that by matching the power laws for the dilute and dense limits at $N/g \approx 1$
it follows that in the semidilute density regime \cite{SMW12}
\begin{equation}
\ninter \approx \frac{\rho a^d}{g^{\nudil \thetadil}} \times (N/g)^{-\nu \thetatwo}
\sim \rho^{21/8} N^{-3/8},
\label{eq_ninter_rho}
\end{equation}
i.e. the fraction of monomers in interchain contact increases strongly with density.

\paragraph*{Focus of present study.}
Having received experimental attention recently \cite{mai00,Deutsch05,Monroy05,Monroy07,Kumaki12},
the aim of the present study is to discuss the density dependence of various thermodynamic properties
such as the pressure $P$ or the compression modulus $K$ of the solution 
\cite{LandauElasticity,RowlinsonBook,HansenBook}
focusing on the experimentally relevant semidilute regime. 
Comparing various numerical techniques we will confirm, e.g., that the dimensionless compressibility
$\gT(\rho) \equiv \kBT \rho/K$ scales as the blob size $g(\rho)$ 
as one expects according to a standard density crossover scaling \cite{DegennesBook}.
This is of some importance since due to the generalized Porod scattering of the compact chains
the intrachain coherent structure factor $F(q)$ has been shown to scale as 
$N F(q) \approx N^2 / (q R)^{2d-\dperi}$ 
in the intermediate regime $1/R \ll q \ll 1/\xi$ of the wavevector $q$  \cite{MKA09,MWK10,MSZ11,SMW12}. 
It is hence dangerous to determine the blob size by means of an Ornstein-Zernike fit as done, 
e.g., in ref.~\cite{mai00}.

\paragraph*{Outline.}
We begin the discussion in sect.~\ref{sec_algo} by summarizing our coarse-grained model 
and the computational schemes used. Our numerical results are presented
in sect.~\ref{sec_res}. We remind first in sect.~\ref{res_RNRs} the scaling of the chain 
and subchain size already presented elsewhere \cite{SMW12} and characterize the size 
of the semidilute blob. We present then different energy and pressure contributions,
the dimensionless compressibility $\gT$ (sect.~\ref{res_compress}) and the elastic moduli $\Xvirial$ and $\Xfluctu$
characterizing, respectively, the affine linear response to an external  homogeneous strain and the counteracting stress 
fluctuations (sect.~\ref{res_moduli}) \cite{Lutsko89,WTBL02,SBM11,paptrunc}.
Some complementary information concerning the elastic Lam\'e coefficients $\lambda$ and $\mu$ \cite{LandauElasticity}
and their various contributions is referred to the Appendix. 
How the blob size may be determined in a real experiment
using the scaling of the total structure factor $S(q)$ is shown in sect.~\ref{res_Sq}.
We conclude the paper in sect.~\ref{sec_conc} where we comment on the relevance of our findings
for polymer blends confined to ultrathin slits and the influence of a finite persistence length.


\section{Coarse-grained polymer model and computational issues}
\label{sec_algo}

\paragraph*{Effective Hamiltonian.}
The aim of the present study is to clarify universal power-law scaling predictions in the limit of 
large chain length $N$ and low wavevector $\qvec$ where the specific physics and chemistry on monomeric 
level is only relevant for prefactor effects \cite{DegennesBook,BWM04}. 
As in our previous studies \cite{MKA09,MWK10,MSZ11,SMW12,WMJ10} we sample solutions and melts of monodisperse, 
linear and highly flexible chains using a version of the well-known Kremer-Grest (KG)
bead-spring model \cite{KG86,LAMMPS}.
The non-bonded excluded volume interactions between the effective monomers are represented by a purely 
repulsive (truncated and shifted) Lennard-Jones (LJ) potential
\cite{AllenTildesleyBook,FrenkelSmitBook}
\begin{equation}
u_{\text{nb}}(r) = 4 \epsilon \left[ (\sigma/r)^{12} - (\sigma/r)^6 \right] + \epsilon 
\ \mbox{ for } r /\sigma \le 2^{1/6} 
\label{eq_algoLJ}
\end{equation}
and $u_{\text{nb}}(r) = 0$ elsewhere \cite{foot_truncated}. 
At variance to the standard KG model {\em (i)} the LJ potential is assumed {\em not} act between adjacent 
monomers \cite{foot_connectedmons} and 
{\em (ii)} these bonded monomers are connected by a simple harmonic spring potential \cite{foot_harmonic}
\begin{equation}
\ubond(r) = \frac{1}{2} \kbond (r - \lbond)^2
\label{eq_algoSPRING}
\end{equation}
with a spring constant $\kbond = 676 \epsilon$ and a bond reference length $\lbond = 0.967 \sigma$ 
calibrated to the ``finitely extendible nonlinear elastic" (FENE) springs of the original KG model 
\cite{KG86}. 
No additional stiffness term has been included and, at variance to most experimental systems 
\cite{mai00,Deutsch05,Rubinstein07,Monroy11}, 
our chains are flexible down to monomeric scales. Effects of finite persistence length are only briefly alluded
to in the outlook presented at the end of the paper.

\paragraph*{Units and non-intersection constraint.}
The monomer mass $m$, the temperature $T$ and Boltzmann's constant $\kB$ are all set to unity, 
i.e.  $\beta \equiv 1/\kBT = 1$ for the inverse temperature, and 
LJ units ($\epsilon=\sigma=m=1$) are used throughout the paper.
The parameters and settings of our model make chain intersections impossible. 
(It has been explicitly checked that such a violation never occurs.) 
This is illustrated in the snapshot presented in fig.~\ref{fig_snap_semi} for chains in 
the semidilute density regime at number density $\rho=0.125$, chain length $N=2048$,
chain number $M=48$ and linear box dimension $\Lbox \approx 887$. 
We simulate thus strictly 2D self-avoiding walks as required.
\begin{figure}[t]
\centerline{\resizebox{0.8\columnwidth}{!}{\includegraphics*{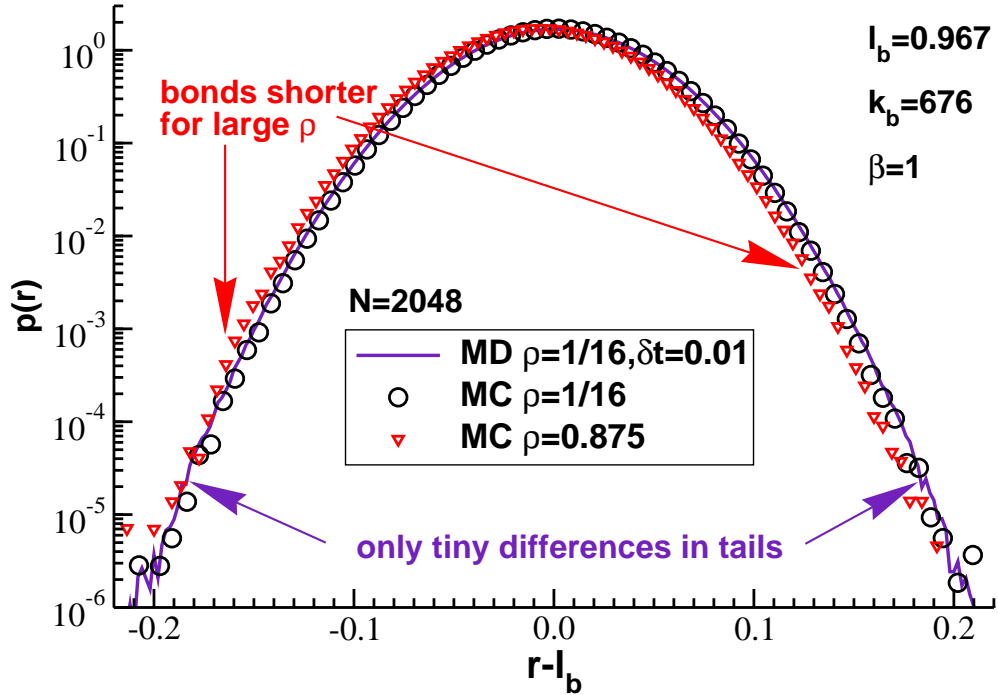}}}
\caption{Normalized radial bond length distribution $p(r)$ {\em vs.} $r-\lbond$ for $N=2048$.
For all ensembles we obtain essentially $p(r) \propto \exp(- \beta \ubond(r))$ 
as expected if the excluded volume interactions were rigorously switched off. 
Note that for MD simulations with $\delta t =0.01$ the tails of $p(r)$
are slightly larger at low densities (solid line) compared to the MC simulations (circles)
as shown for $\rho=0.0625$ where $l = 0.9700$. 
At larger densities the excluded volume interactions reduce the distance between
bonded beads as shown for $\rho=0.875$ (triangles) where $l = 0.963$.
\label{fig_bondhisto}
}
\end{figure}

\paragraph*{Configuration sampling.}
Taking advantage of the public domain LAMMPS implementation \cite{LAMMPS}
the bulk of the presented results for the more time consuming higher densities \cite{foot_comppower}
has been obtained by MD simulation integrating the classical equations of motion with 
the Velocity-Verlet algorithm \cite{AllenTildesleyBook,FrenkelSmitBook} 
using the standard time step $\delta t = 0.01$ \cite{KG86,MKA09,MWK10,MSZ11,SMW12,WMJ10}.
The constant temperature $T=1$ is imposed by means of a Langevin thermostat 
\cite{AllenTildesleyBook,FrenkelSmitBook,KG86} with a friction constant $\gamma=0.5$.
Note that the Langevin thermostat is believed to stabilize the integration at a larger 
time step (an improvement of about a factor $10$ is reported) 
than necessary for microcanonical simulations \cite{Duenweg97a}.
We emphasize that the strong harmonic bonding potential --- used to avoid chain intersections --- corresponds 
to a small oscillation time $\Tbond = 2\pi \sqrt{ m/\kbond} \approx 0.24$.
Since $\Tbond$ is only an order of magnitude larger than $\delta t$,
this begs the question of whether configurations of correct statistical weight have been sampled.

In order to crosscheck our results we have in addition performed Monte Carlo (MC) simulations which (by construction)
obey detailed balance \cite{FrenkelSmitBook}, i.e. produce an ensemble of configurations with correct weights. 
A mix of local monomer moves (with displacement attempts uniformly distributed in a disk of radius $\delta r = 0.1$) 
and global slithering snake moves along the chain contours is used \cite{BWM04}. The latter slithering snake 
moves turn out to be efficient for exploring the configuration space at low densities up to 
$\rho \approx 0.25$. As one expects, for larger densities the acceptance rate of 
the snake moves deteriorates, since it becomes too unlikely to find enough free volume to place a new chain end. 

The comparison of ensembles generated with both methods shows that all sampled properties 
are essentially identical. This can be seen, e.g., in fig.~\ref{fig_bondhisto} for the bond length distribution $p(r)$ for 
chains of length $N=2048$. Note that  $\int \ddiff r (2\pi r) p(r) =1$ and $\int \ddiff r (2\pi r) p(r) r^2 =l^2$.
The expected weak density effect for the bond length is clearly seen in the figure.   
See Table~\ref{tab_rho} for the root-mean square bond length $l(\rho)$.
Only closer inspection reveals that the MD method with $\delta t = 0.01$ yields slightly too large $l$
at lower densities and the tails of $p(r)$ are weakly enhanced when compared to our MC results. 
We shall see in sect.~\ref{res_pressure} that this tiny numerical effect matters for the computation of the pressure 
in the dilute density regime.

\paragraph*{Parameter range.}
Some relevant conformational and thermodynamic properties are summarized in Table~\ref{tab_rho} 
for our main reference chain length $N=1024$. As in ref.~\cite{SMW12} where we have discussed various 
conformational properties of semidilute solutions and melts we scan over a broad range of densities $\rho$. 
All properties reported for $\rho \le 0.25$ have been sampled using MC \cite{foot_comppower}.
The highest density we have computed is $\rho=1.0$ for a chain length $N=64$.
In cases where chain length does not matter this data set is often presented together 
with data obtained for $N=1024$ at lower densities.
Note that our largest chain length $N=2048$ is about an order of magnitude larger than 
in previous computational studies of semidilute solutions and melts in two dimensions
\cite{carmesin90,Rutledge97,Yethiraj03,Cavallo03,Cavallo05}.
To avoid finite system size effects the periodic simulation boxes contain at least $M=96$ chains for the higher densities.
Even more chains are sampled for shorter chains.

\section{Numerical results}
\label{sec_res}

\subsection{Chain and subchain size}
\label{res_RNRs}

\paragraph*{Dilute reference limit.}
A density crossover scaling study requires the precise characterization of the dilute limit \cite{DegennesBook}. 
Using slithering snake MC moves we have thus sampled single chain systems ($M=1)$ with chain lengths up to $N=8192$. 
To avoid the self-interaction of the chains with their periodic images huge simulation boxes 
($\Lbox = 10^4$, $\rho \approx 0$) are used. As one expects \cite{DegennesBook}, 
the typical chain size is seen to increase with a power-law exponent $\nu=\nudil \equiv 3/4$ (not shown).
For the root-mean-square chain end-to-end distance $\Rend(N)$ we obtain $\Rend(N)/N^{\nudil} \to \benddil \approx 0.98$ 
for the largest chain lengths probed. The corresponding effective segment size for the
dilute radius of gyration $\Rgyr(N)$ is found to become
\begin{equation}
\lim_{N \to \infty} \Rgyr(N)/N^{\nudil} \equiv \bgyrdil \approx 0.37.
\label{eq_bdil_def}
\end{equation}
Dropping the index ``{\em g}" the latter length scale $\bdil$ is used below
to make various properties dimensionless, allowing thus a meaningful comparison with
experiments or other computational models.

\begin{figure}[t]
\centerline{\resizebox{.8\columnwidth}{!}{\includegraphics*{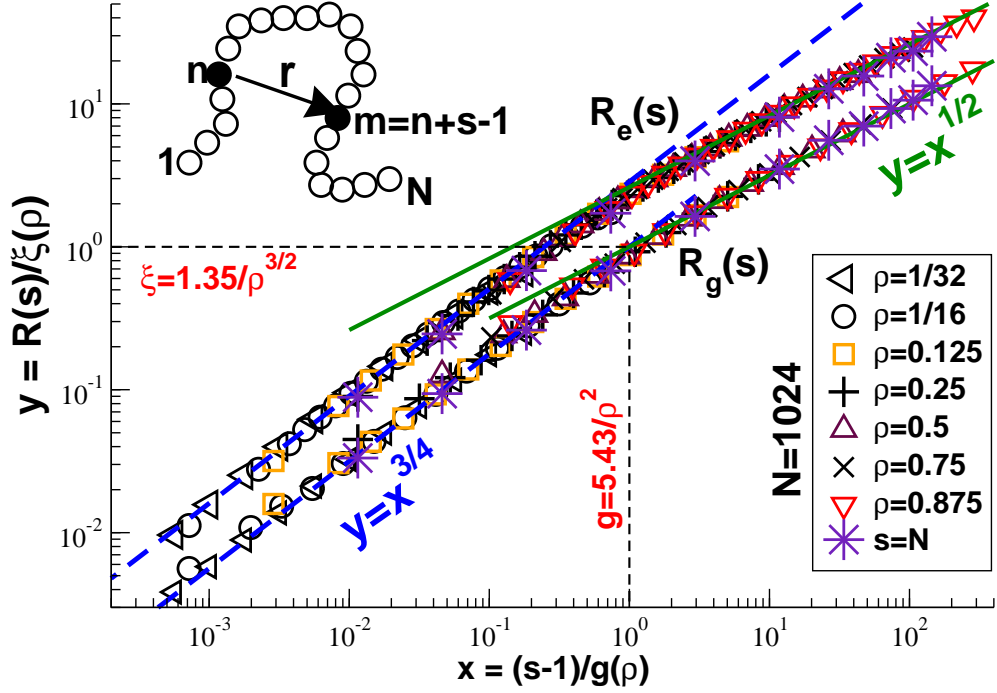}}}
\caption{Root-mean-square end-to-end distance $\Rend(s)$ and radius of gyration 
$\Rgyr(s)$ for subchains of $s = m - n+1 \le N$ monomers with $N=1024$.  
The total chain end-to-end distance $\Rend(N)$ and radius of gyration $\Rgyr(N)$ 
are indicated by stars ($s=N$).
Plotting $\Rend(s)/\xi(\rho)$ and $\Rgyr(s)/\xi(\rho)$ {\em vs.} $(s-1)/g(\rho)$ yields
a perfect data collapse for all densities. 
\label{fig_RNRs}
}
\end{figure}

\paragraph*{Scaling for finite densities.}
That sufficiently long 2D polymer chains become indeed compact for all densities, 
as stated by eq.~(\ref{eq_compact}), is reminded in fig.~\ref{fig_RNRs}.
We present here the root-mean-square end-to-end distance $\Rend(s)$ and the radius of gyration $\Rgyr(s)$
of subchains of length $s = |m-n+1| \le N = 1024$ as sketched in the figure. 
The averages are taken over all pairs of monomers $(n,m=n+s-1)$ possible.
Averaging only over subchains at the curvilinear chain center ($n,m \approx N/2$)
slightly reduces chain end effects; however, the difference is negligible for the large chains we focus on.
The limit $s=N$ corresponds obviously to the total chain size which is represented in the figure
for chains of length $N=1024$ (stars).
As expected, the typical (sub)chain size increases with an exponent $\nu=\nudil \equiv 3/4$ 
in the dilute limit (dashed lines) and with $\nu=1/d$ for larger (sub)chains and densities
($x \gg 1$) in agreement with various numerical \cite{carmesin90,Rutledge97,Yethiraj03} and 
experimental studies \cite{mai00,Rubinstein07}. 
The subchain size is represented here to remind that not only the total chain 
becomes compact but in a self-similar manner the chain conformation 
on all scales \cite{MWK10}.

\paragraph*{Operational definition of blob size.}
In agreement with the standard density crossover scaling, eq.~(\ref{eq_grho}), the axes
of fig.~\ref{fig_RNRs} have been made dimensionless by plotting $y = R(s)/\xi(\rho)$ as 
a function of $x = (s-1)/g(\rho)$ where we define
\begin{eqnarray}
\xi & \equiv & \bdil g^{\nudil} \approx 1.35/\rho^{3/2} \mbox{ and } \nonumber \\
g & \equiv & 0.09/(\bdil^d \rho)^{1/(\nudil d-1)}  \approx 5.43/\rho^2. 
\label{eq_blobsize}
\end{eqnarray}
The (slightly arbitrary) prefactors have been fitted using $\Rgyr(s)$ for $N=1024$ and $N=2048$ for 
the densities $\rho=0.0625$ and $\rho=0.125$.
This yields a perfect data collapse especially considering that a broad range of densities is considered. 
Please note that the asymptotic power-law slopes for $x \ll 1$ (dashed lines) and $x \gg 1$ (bold lines) for 
the radius of gyration intersect exactly at $(x,y)=(1,1)$.
Since $\Rend(s)$ becomes compact more rapidly than $\Rgyr(s)$, a blob size defined using $\Rend(s)$ would be slightly smaller. 

\paragraph*{Scaling for large densities.}
Interestingly, the blob scaling of the (sub)chain size is even successful for our highest densities
where the blob picture clearly breaks down for some other properties as discussed, e.g.,
in sect.~\ref{res_pressure} below. This is due to the fact that $R(s)$ is set in this limit by the
typical distance $\dcm \approx (s/\rho)^{1/d}$ between the chain or subchain centers of mass and
this irrespective of the physics (monomer size, persistence length, $\ldots$) on small scales.
Since $\rho \approx g/\xi^d$ is imposed, a different choice of the blob size only leads to a {\em shift}
of the data along the bold power-law slope $\nu=1/d$. Only for sufficiently low densities where
(sub)chains smaller than the blob can be probed it is possible to test the blob scaling and to
adjust the prefactors.

\begin{figure}[t]
\centerline{\resizebox{0.8\columnwidth}{!}{\includegraphics*{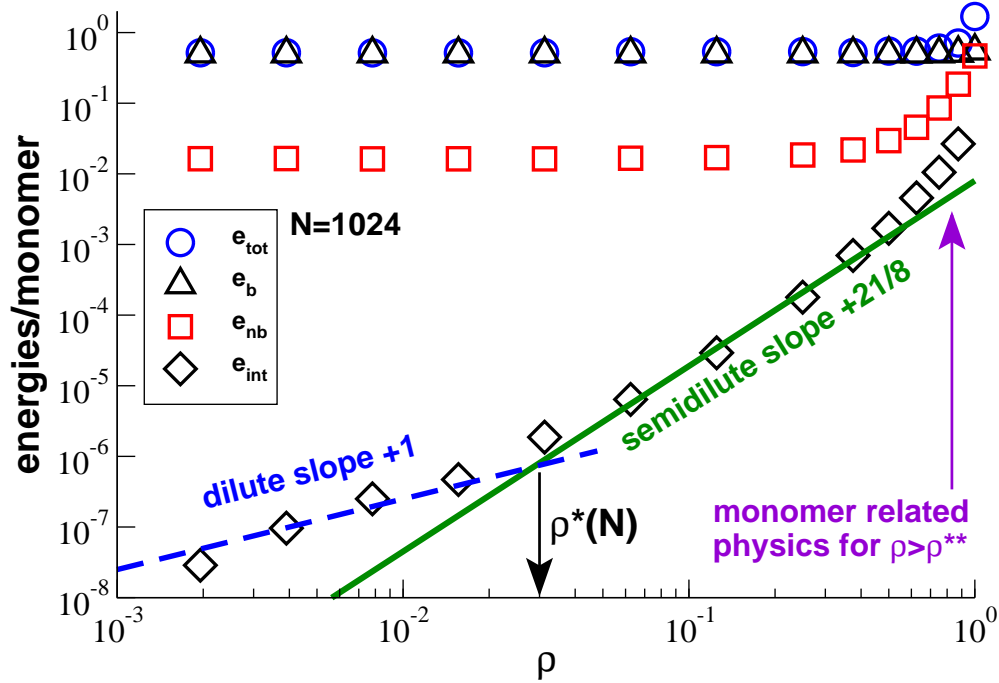}}}
\caption{Various mean energy contributions per monomer {\em vs.} density $\rho$
for chain length $N=1024$: 
Total potential energy $\etotal$, bonding energy $\ebond$, excluded volume energy $\enb$ 
and interaction energy $\einter$ between monomers of different chains.
The power-law slope predicted for the semidilute regime is indicated by the bold line.
The vertical arrow indicates $\rhostar$ for $N=1024$ as determined from the pressure isotherm,
eq.~(\ref{eq_rhostarprefactor}), discussed below.
Non-universal behavior due to the specific monomer interactions is seen at higher densities 
with $\rho > \rhostarstar \approx 0.5$.
\label{fig_energy_rho}
}
\end{figure}

\begin{figure}[t]
\centerline{\resizebox{0.8\columnwidth}{!}{\includegraphics*{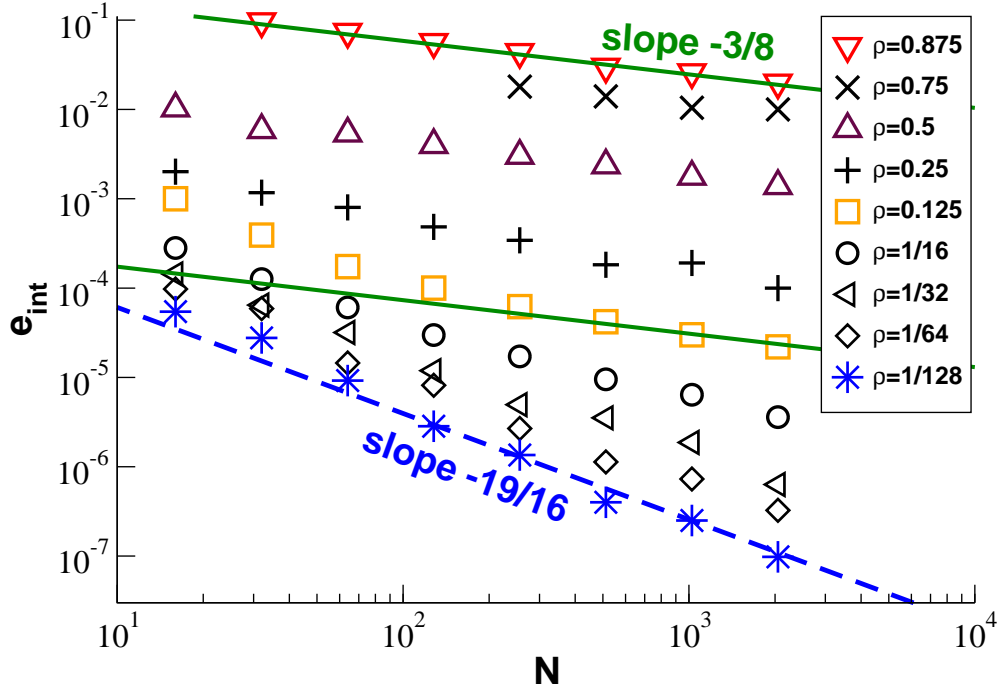}}}
\caption{Interchain interaction energy per monomer $\einter$ {\em vs.} chain length $N$
for different $\rho$. The dashed and solid power-law slopes with $-\nu \thetatwo = - 19/16$
and $-3/8$ represent, respectively, the expected asymptotic behavior for dilute and dense solutions. 
\label{fig_energy_N}
}
\end{figure}

\subsection{Energy}
\label{res_energy}

From the numerical point of view the simplest thermodynamic property to be investigated
here is the total mean potential energy per monomer $\etotal$ due to the Hamiltonian 
described in sect.~\ref{sec_algo}. As shown in fig.~\ref{fig_energy_rho}, 
it is essentially density independent and always dominated by the bonding potential $\ebond$.
Due to the harmonic springs used we have $\ebond \approx \kBT/2$ as expected according to the 
equipartition theorem \cite{foot_harmonic}. Noticeable (albeit weak) corrections to this value are only found
for our highest densities where bonded monomer pairs are pushed together by the excluded volume interactions
as already seen in fig.~\ref{fig_bondhisto}.
The total non-bonded excluded volume interaction per monomer $\enb$ 
becomes constant at low densities where it is dominated by the excluded volume
interaction of curvilinear neighbors on the {\em same} chain. The non-bonded energy $\enb$
increases for larger densities, but remains always smaller than the
bonded energy $\ebond$. Values of $\enb$ for $N=1024$ are included in Table~\ref{tab_rho}.

From the theoretical point of view more interesting is the contribution to the
total excluded volume interaction due to the contact of monomers from different 
chains measured by $\einter$. For not too high densities $\einter$
is expected to scale as the fraction $\ninter$ of monomers in interchain contact
mentioned in the Introduction. 
As indicated by the dashed line the interchain energy $\einter$ is 
proportional to the density in the dilute regime (dashed line) due to the 
mean-field probability that two chains are in contact. At higher semidilute
densities above the crossover density $\rhostar(N)$ 
up to $\rhostarstar \approx 0.5 N^0$ a much stronger power-law exponent $\approx 21/8$ is seen
in agreement with the established scaling for $\ninter$, eq.~(\ref{eq_ninter_rho}).
The observed $\rho$-dependence of $\einter$ in the semidilute regime is thus traced back to the known
values of the universal exponents $\nu$ and $\thetatwo$ in the dilute and dense limits.
The interaction energy $\einter$ is seen to increase even more strongly for densities $\rho > \rhostarstar$
where the semidilute blob picture becomes inaccurate.
At variance to the other mean energies,  $\einter$ has a strong chain length effect as revealed 
in fig.~\ref{fig_energy_N}. The indicated power-law slopes correspond to the expected exponents 
$\nu \thetatwo = 19/16$ and $3/8$ for, respectively, the dilute (dashed line) and dense (bold lines) 
density limits. Since the $N$-scaling does not require the existence of sufficient large semidilute blobs,
it even holds for our highest ``melt" densities.


\subsection{Pressure}
\label{res_pressure}

\paragraph*{Definitions and virial equation.}
While the energy contributions discussed above cannot be probed in a real experiment,
the osmotic pressure of 2D polymer systems can be accessed experimentally 
for polymer solutions at the air-water interface \cite{Deutsch05,Monroy05,Monroy07,Monroy10,Monroy11,Kumaki12}.
As usual for pairwise additive interactions the mean pressure $P = \Pid + \Pex$ is
obtained in our simulations as the sum of the ideal kinetic contribution $\Pid = \kBT \rho$
and the excess pressure contribution 
\begin{equation}
\Pex = \Pnb + \Pbond = \la \Wvirial \ra/V.
\label{eq_Pex}
\end{equation}
Here, $\Pnb$ stands for the non-bonded excess pressure contribution, $\Pbond$ for the bonded pressure contribution,
$V$ for the $d$-dimensional volume, i.e. the surface $\Lbox^2$ of our periodic simulation box,
and $\Wvirial$ for the internal virial \cite{AllenTildesleyBook,FrenkelSmitBook}
\begin{equation}
\Wvirial = \frac{1}{d} \sum_{l} \rijl \fijl = -\frac{1}{d} \sum_{l} w(\rijl) 
\label{eq_Pdefine}
\end{equation}
with $\fijl$ being the force of the interaction $l$ between two beads $i$ and $j$ at a 
distance $\rijl \equiv ||\rvec_i-\rvec_j||$ 
and $w(\rijl) = \rijl u^{\prime}(\rijl)$ the virial function associated with the 
bonded and non-bonded pair potential $u(\rijl)$. 
(The sum $\sum_l$ stands for the double sum $\sum_{i < j}$ over all monomers $\nmon$ of the solution.)
\begin{figure}[t]
\centerline{\resizebox{.8\columnwidth}{!}{\includegraphics*{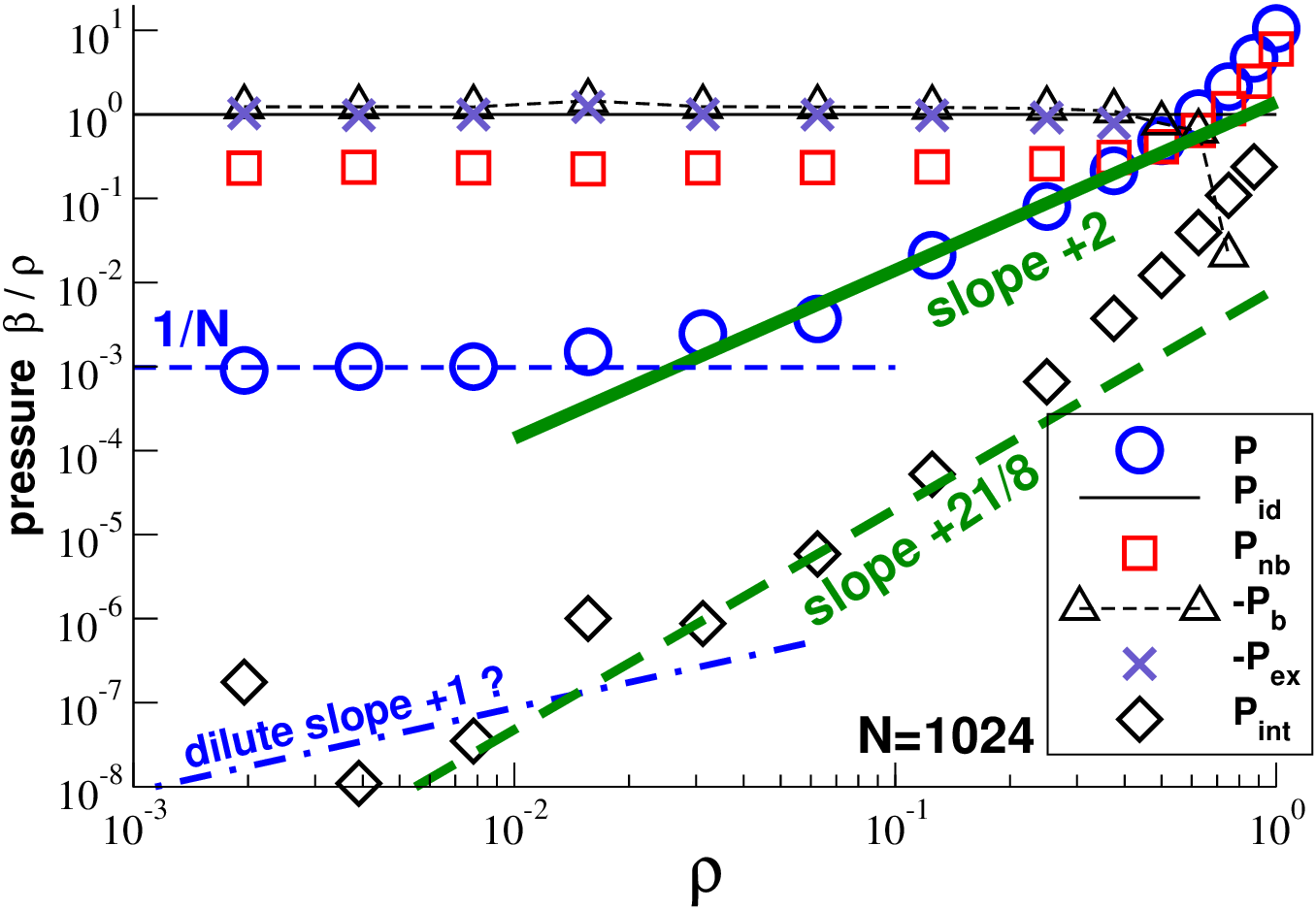}}}
\caption{Various contributions to the isothermal pressure for chains of length $N=1024$
as a function of density $\rho$:
the total pressure $P = \Pid + \Pex$,
the ideal kinetic pressure $\Pid = \kBT \rho$,
the (negative) bonded pressure contribution $-\Pbond$,
the non-bonded LJ interaction contribution $\Pnb$,
the (negative) excess pressure $-\Pex=-(\Pbond+\Pnb)$ and 
the pressure contribution $\Pinter$ due to the LJ interactions between monomers on different chains.
The thin dashed line indicates the dilute total pressure limit $P \beta/\rho = 1/N$,
the bold line the exponent $d\nudil/(d\nudil-1) -1 = 2$ according to eq.~(\ref{eq_Psemidilute})
and the dashed bold line the exponent expected for $\Pinter$ in the semidilute regime.
\label{fig_pressure_rho}
}
\end{figure}

\paragraph*{Pressure contributions.}
We present in fig.~\ref{fig_pressure_rho} the different contributions to the total pressure 
$P = \Pid + \Pnb + \Pbond$ as a function of density $\rho$ focusing on the chain length $N=1024$. 
The vertical axis is rescaled by a factor $\beta/\rho$ to present the contributions {\em per monomer} 
as in fig.~\ref{fig_energy_rho} for the different energy contributions. 
The dilute limit, where $P \beta/\rho \to 1/N$, is indicated by the thin dashed line, 
the semidilute limit by the bold line representing the expected power law $P  \beta/\rho \sim \rho^2$
discussed below.
The rescaled non-bonded pressure $\Pnb$ becomes constant for densities $\rho < \rhostarstar \approx 0.5$
with a plateau value $\Pnb \beta/\rho \approx 0.23$ due to the intrachain interactions of closely 
connected neighbors along the chain as the corresponding energy contribution $\enb$ presented in fig.~\ref{fig_energy_rho}.

In the same density regime we have $\Pbond \beta/\rho \approx -1.23$ for the pressure contribution $\Pbond$ due 
to the bonding potential. 
We remind that for asymptotically long non-interacting phantom chains 
one would have $\Pbond \beta/\rho = -1$ irrespective of the details of the bonding potential \cite{foot_Pphantom}.
At variance to the energy contribution $\ebond$ the corresponding pressure contribution $\Pbond$ is thus
slightly changed by the presence of the excluded volume potential.
The pressure contribution $\Pbond$ must be more negative, of course, since the excess pressure 
$\Pex = \Pnb + \Pbond$ must essentially cancel (for very large chains) the ideal pressure $\Pid$ 
indicated by the thin solid line.  
Note also that for our highest densities where the bonded monomers are pressed together both $\Pbond$ and $\Pex$ become positive,
i.e. are not represented in the figure.

The pressure contribution $\Pinter \beta/\rho$ due to interchain monomer contacts (squares)
shows, not surprisingly, the same power-law exponents as in the interchain interaction energy $\einter$
discussed above. In the semidilute regime we confirm $\Pinter \beta/\rho \sim \rho^{21/8}$ as indicated
by the bold dashed line.
The expected dilute limit $\Pinter \beta/\rho \sim \rho$ (dash-dotted line) 
is unfortunately not yet confirmed numerically due to insufficient statistics
and the error bars (not given) become much larger than the symbol size \cite{foot_PinterN}.
\begin{figure}[t]
\centerline{\resizebox{.8\columnwidth}{!}{\includegraphics*{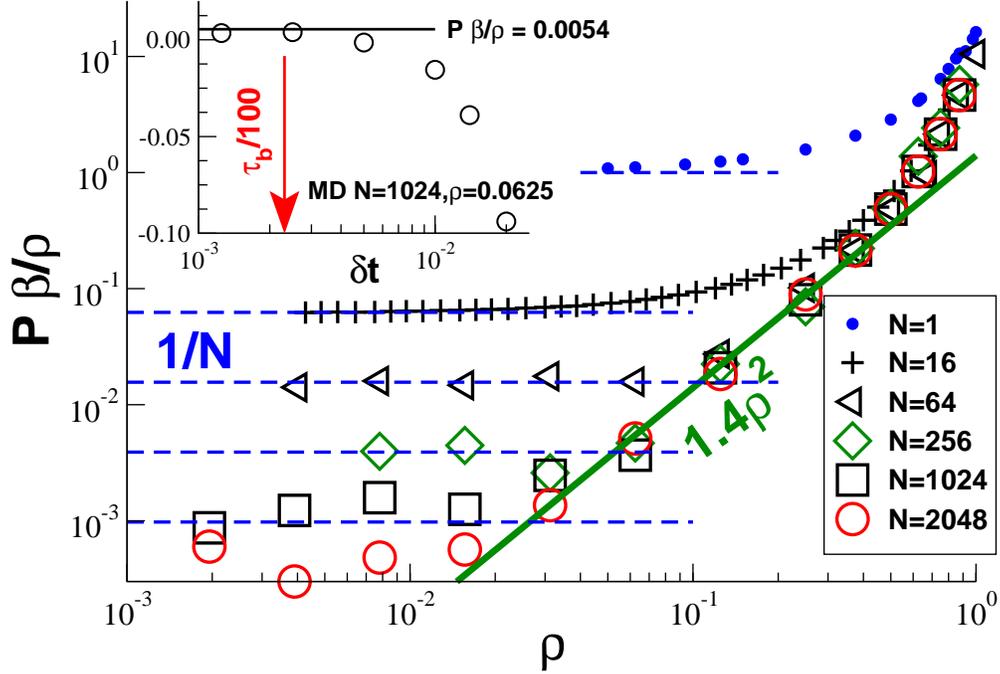}}}
\caption{Total pressure $P \beta/\rho$ as a function of density $\rho$ for different chain lengths $N$. 
Unconnected LJ beads ($N=1$) are indicated by the small filled circles. 
The data for $N=16$ is taken from the instantaneous pressure obtained 
by MD simulation increasing extremely slowly the box size.
The computation of larger chain lengths is much more delicate.
The thin dashed lines give the osmotic pressure in the dilute limit for $N=1$, $16$, $64$, $256$ and $1024$
(from top to bottom).
The pressure becomes $N$-independent with increasing $N$ and $\rho$.
The statistics deteriorates for small $\rho$ and large $N$.
The bold line indicates the power-law exponent $d\nudil/(d\nudil-1)= 3$ expected in 
the semidilute regime according to eq.~(\ref{eq_Psemidilute}).
Inset: 
Pressure as a function of the MD time step $\delta t$ for $N=1024$ and $\rho=0.0625$.
The horizontal line corresponds to our MC result.
As indicated by the vertical arrow the correct pressure 
in the dilute limit is obtained only if $\delta t \ll \Tbond/100$. 
\label{fig_pressure_scal}
}
\end{figure}

\paragraph*{Scaling with chain length.}
The total pressure $P$ is presented in fig.~\ref{fig_pressure_scal} for a broad range of chain lengths $N$. 
Obviously, $P$  increases strongly with density $\rho$ and the chain length only matters in the dilute limit 
for short chains or small densities where ultimately the translational entropy of the chains governs the free energy 
\cite{DegennesBook} as indicated by the thin dashed horizontal lines.
This (theoretically trivial) limit turns out to be numerically challenging
since a large positive term, the ideal pressure contribution $\Pid$, is nearly canceled by a large negative term, 
the excess pressure $\Pex \approx \Pbond$, which requires increasingly good statistics as $N$ becomes larger. 
The precise determination of $P$ becomes surprisingly time consuming even if the configurations are 
sampled by means of slithering snake MC moves (sect.~\ref{sec_algo}). Given sufficient numerical precision 
this yields, however, the expected dilute pressure as is seen from the main panel of the figure.
This is different if systems are computed using MD with the standard time step $\delta t = 0.01$ 
as shown in the inset of fig.~\ref{fig_pressure_scal}. 
The bond oscillation time $\Tbond/100$ is indicated by the vertical arrow. 
As we have already seen in fig.~\ref{fig_bondhisto}, the bonds become
slightly stretched if $\delta t \gg \Tbond/100$, i.e. they become too tensile 
which corresponds to a too negative $\Pbond$. Unfortunately, the bonding potential is that strong,
i.e. $\Tbond/100$ so small, that it gets too time consuming to compute the correct pressure in this density
limit using MD \cite{foot_Pbest}.

\paragraph*{Semidilute density regime.}
Of experimental interest beyond these computational issues is the intermediate semidilute density regime indicated
by the bold power-law slope. The universal exponent can be understood by an elegant crossover scaling 
argument given by de Gennes \cite{DegennesBook} where the pressure is written as 
$\beta P = \rho/N \times f(\rho/\rhostar)$ with $\rhostar(N) \approx N/R^d(N) \approx N^{1-d\nudil}/ \bdil^d$ 
being the crossover density and $f(x)$ a universal function \cite{foot_Pscal}. 
Assuming $P$ to be chain length independent for $x \gg 1$
this implies $f(x) \sim x^{1/(d\nudil-1)}$ and, hence,
\begin{equation}
\beta P \bdil^2  \approx (\bdil/\xi(\rho))^d \approx (\bdil^d \rho)^{d\nudil/(d\nudil-1)} \approx (\bdil^2 \rho)^3
\label{eq_Psemidilute}
\end{equation}
where we have used eq.~(\ref{eq_grho}) to restate the well-known relation between 
pressure and blob size $\xi$  \cite{DegennesBook}. 
The predicted exponent fits the data over about a decade in density where the blob size is sufficiently large. 
Additional physics becomes relevant for densities around $\rhostarstar$ 
where the LJ excluded volume starts to dominate all interactions, i.e. the pressure of polymer chains 
approaches the pressure of unbonded LJ beads (filled circles).

\paragraph*{Universal amplitude.}
The choice of the axes of the main panel of fig.~\ref{fig_pressure_scal} 
may make the comparison to real experiments difficult. Choosing as a (natural but arbitrary)
length scale the effective segment size $\bdil$ associated to the dilute radius of gyration,
eq.~(\ref{eq_bdil_def}), one may instead plot the rescaled pressure $y = \beta P \bdil^d$ 
as a function of the reduced density $x = \bdil^d \rho$.  
In the semidilute regime this corresponds to a power-law slope $y = \aP x^3$ with a power-law 
amplitude $\aP \approx 83.4$. 
The dimensionless amplitude $\aP$ (or similar related values due to different choices of $\bdil$) 
should be compared to real experiments or other computational models. 
As long as the blob size is sufficiently large, i.e. $\rho$ small enough,
molecular details should not alter this universal amplitude. 
Persistence length effects, e.g., change $\bdil$ but not $\aP$.
Using eq.~(\ref{eq_blobsize}) this implies 
\begin{equation}
\beta P \approx  0.16 \aP /\xi^2 \approx 13.7/\xi^2
\label{eq_P2xi}
\end{equation}
with universal prefactors (within the operational definition based on the radius of gyration)
allowing thus to determine the blob size from the experimentally obtained pressure isotherms.
We note finally that by matching the dilute asymptote with the 
semidilute pressure regime the prefactor of the crossover density $\rhostar$
may be operationally defined as
\begin{equation}
\rhostar \equiv \aP^{1-d\nudil} \times N/\Rgyr^d \approx 0.845/N^{1/2} 
\label{eq_rhostarprefactor}
\end{equation}
with $\Rgyr$ being the radius of gyration in the dilute limit. This implies $\rhostar \approx 0.02$
for our largest chains with $N=2048$. Considering that the semidilute regime breaks down at $\rho \approx \rhostarstar$
this limits the semidilute scaling to about an order of magnitude in density.

\begin{figure}[t]
\centerline{\resizebox{.8\columnwidth}{!}{\includegraphics*{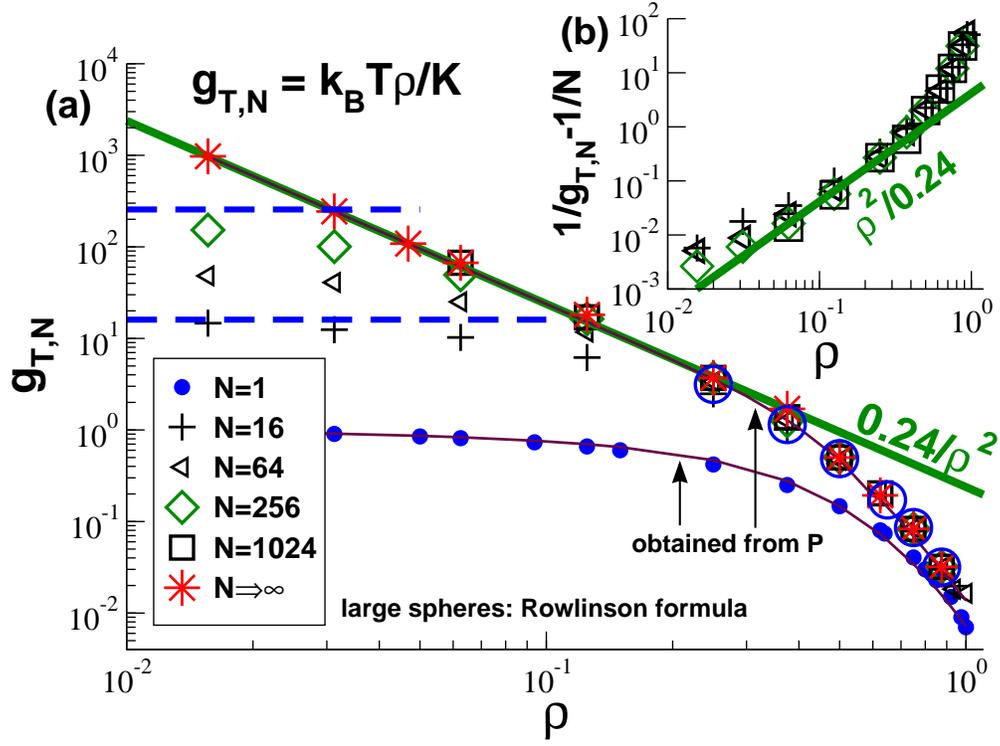}}}
\caption{Determination of dimensionless compressibility $\gT(\rho) = \lim_{N\to \infty} \gTN(\rho)$:
{\bf (a)} $\gTN(\rho)$ obtained by various means for several chain lengths $N$.
The thin lines correspond to a polynomial fit to $P(\rho)$ for $N=1$ (lower thin line)
and the largest chains available (upper thin line). The large circles have been obtained
using the Rowlinson formula, eq.~(\ref{eq_KRowlinson}), for $N=1024$. 
All other data correspond to the plateau of the total structure factor $S(q,N)$ for small wavevectors $q$.
The bold line indicates eq.~(\ref{eq_grho}),
the two horizontal dashed lines the dilute limit for $N=16$ and $N=256$. 
{\bf (b)} Data collapse of $1/\gTN - 1/N$ for different $N$ with $\gTN$ obtained from the
total structure factor. 
\label{fig_compress}
}
\end{figure}


\subsection{Compressibility}
\label{res_compress}

\paragraph*{Definitions.}
Being an isotropic liquid, a polymer solution is described in the hydrodynamic limit
by only {\em one} experimentally relevant elastic modulus, the bulk compression modulus 
\cite{HansenBook,RowlinsonBook} 
\begin{equation}
K \equiv 1/\kappaT \equiv \kBT \rho / \gTN =  \rho \frac{\partial P}{\partial \rho},
\label{eq_Kmodulus}
\end{equation}
with $\kappaT$ being the standard isothermal compressibility 
and $\gTN$ the ``dimensionless compressibility" for systems of finite chain length $N$.
We use here the additional index $N$ to distinguish $\gTN$ from the
dimensionless compressibility for asymptotically long chains $\gT \equiv \lim_{N\to \infty} \gTN$.
Due to the translational entropy of the chains (van't Hoff's law) both quantities are related by 
\cite{DegennesBook,WCX11}
\begin{equation}
\frac{1}{\gTN} = \frac{1}{N} + \frac{1}{\gT},
\label{eq_gTNgT}
\end{equation}
i.e. $\gTN$ and $\gT$ are expected to differ strongly for small densities where $\gT$ must be large.
Our aim is to determine $\gTN$ precisely comparing different techniques and to extrapolate then
using eq.~(\ref{eq_gTNgT}) to $\gT$ which should scale in the semidilute regime as the number of 
monomers per blob, eq.~(\ref{eq_grho}). 

\paragraph*{Measurements.}
Using eq.~(\ref{eq_Kmodulus}) the bulk modulus and/or the dimensionless compressibility can be 
obtained, of course, from the pressure isotherms discussed in the previous subsection. 
This is best done by fitting a spline to $y \equiv \log(\beta P)$ as a function of $x \equiv \log(\rho)$.
The resulting curves for $N=1$ and the largest chain lengths available are 
represented by the thin lines in the main panel of fig.~\ref{fig_compress} 
where $\gTN$ is traced as a function of density $\rho$. 
A disadvantage of this method is of course that $P(\rho,N)$ must be known for a large number 
of densities, especially for large $\rho$ where the pressure increases strongly.

Alternatively, the compressibility for one specific density can be directly computed
using the Rowlinson stress fluctuation formula \cite{RowlinsonBook,AllenTildesleyBook}
\begin{equation}
K =  P + \Xvirial - \Xfluctu
\label{eq_KRowlinson}
\end{equation}
where we emphasize that $K$ depends explicitly on the total pressure $P$.
The second contribution $\Xvirial$ represents the so-called ``hypervirial" 
\begin{equation}
\Xvirial \equiv \frac{1}{d^2V}\sum_{l} 
\la \rijl \frac{\ddiff w(\rijl)}{\ddiff \rijl} \ra
\label{eq_Xvirial}
\end{equation}
where $l$ stands again for a pair of monomers $i < j$.
The index $B$ indicates that this contribution corresponds to the famous 
Born approximation of the elastic moduli of solids assuming {\em affine} displacements under an applied
infinitesimal homogeneous strain \cite{Lutsko89,WTBL02,SBM11}. 
(The used notation becomes transparent from the exact relation, eq.~(\ref{eq_laB2Xvirial}), given in the Appendix.) 
This affine approximation overpredicts the free energy change in general.
The overprediction of the compression modulus is ``corrected" by the excess pressure fluctuation 
\begin{equation}
\Xfluctu \equiv \beta V 
\left(\la (\Wvirial/V)^2 \ra - \la \Wvirial/V \ra^2 \right) \ge 0. \label{eq_Xfluctu}
\end{equation}
Results obtained using eq.~(\ref{eq_KRowlinson}) and local MC moves are indicated by the large circles
in fig.~\ref{fig_compress}. 
While the computation of $K$ using the Rowlinson formula is straightforward for dense systems ($\rho \ge 0.5$),
this becomes for numerical reasons more and more difficult with decreasing density, as further investigated 
in sect.~\ref{res_moduli} \cite{foot_isobaric}.

The bulk of the data presented in fig.~\ref{fig_compress} stems from \cite{HansenBook}
\begin{equation}
\gTN \equiv \rho/ \beta K = \lim_{q \to 0} S(q,N) 
\label{eq_Sq2gT}
\end{equation}
from the plateau in the low-wavevector limit of the total structure factor $S(q,N)$ 
which is further discussed in sect.~\ref{res_Sq}. 
(Equation~(\ref{eq_Sq2gT}) takes advantage of the fact that our monomers are indistinguishable.
It would become much more intricate if the beads were, e.g., polydisperse \cite{HansenBook}.)
Due to our large box sizes 
(especially in the low-$\rho$ limit) this method provides over the whole density 
range reliable numerical values only requiring the analysis of about 1000 more or less
independent configurations.

\paragraph*{Interpretation of data.}
In agreement with eq.~(\ref{eq_Kmodulus}) and the scaling of the pressure discussed above,
the dimensionless compressibility $\gTN$ is seen in fig.~\ref{fig_compress} to be strongly $N$-dependent 
for short chains and small densities where the translational entropy matters.
This $N$-dependence is, however, perfectly characterized by eq.~(\ref{eq_gTNgT}) 
as explicitly verified by the scaling collapse presented in the inset. The scaling being 
successful for even rather small chains, this allows us to determine even using short
chains the large-$N$ limit $\gT$ for all densities (indicated by stars).
Consistent with eq.~(\ref{eq_Psemidilute}) the bold line indicates the power-law asymptote 
\begin{equation}
\gT \approx \frac{1}{3 \aP (\bdil^2 \rho)^2} 
\approx 0.24/\rho^2
\label{eq_gTsemidilute}
\end{equation}
for the semidilute regime. Using eq.~(\ref{eq_blobsize}) this implies $g \approx 23\gT$ 
for the number of monomers per blob which may thus be obtained from the dimensionless compresssibility.
%

\subsection{Stress fluctuations}
\label{res_moduli}

\begin{figure}[t]
\centerline{\resizebox{.8\columnwidth}{!}{\includegraphics*{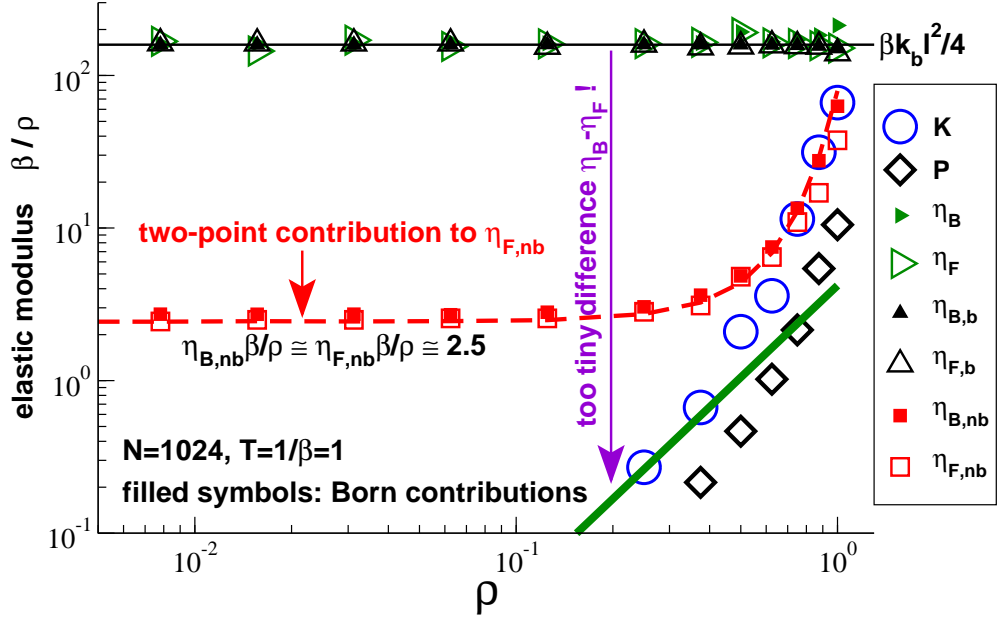}}}
\caption{Contributions to $K$ for $N=1024$ {\em vs.} density $\rho$:
Compression modulus $K = P + \Xvirial - \Xfluctu$, pressure $P$, 
hypervirial $\Xvirial$ and fluctuation contribution $\Xfluctu$.
Also given are the hypervirials $\Xvirialnb$ and $\Xvirialbond$ and
the pressure fluctuation contributions $\Xfluctunb$ and $\Xfluctubond$ associated to the 
non-bonded and the bonded interactions, respectively. 
Filled symbols refer to the hypervirial (Born) contributions.
All data points for $\rho=1.0$ have been obtained for $N=64$.
The horizontal line corresponds to $\Xvirialbond = \beta \kbond l^2/4$ as expected for non-interacting chains,
eq.~(\ref{eq_Xvirialbond}), and the bold solid line to the compression modulus in the semidilute regime
obtained using eq.~(\ref{eq_gTsemidilute}).
The contribution $\Xfluctuselfnb$ associated to the two-point or self correlations of the 
non-bonded interactions is represented by the dashed line.
\label{fig_moduli_rho}
}
\end{figure}

\begin{figure}[t]
\centerline{\resizebox{.8\columnwidth}{!}{\includegraphics*{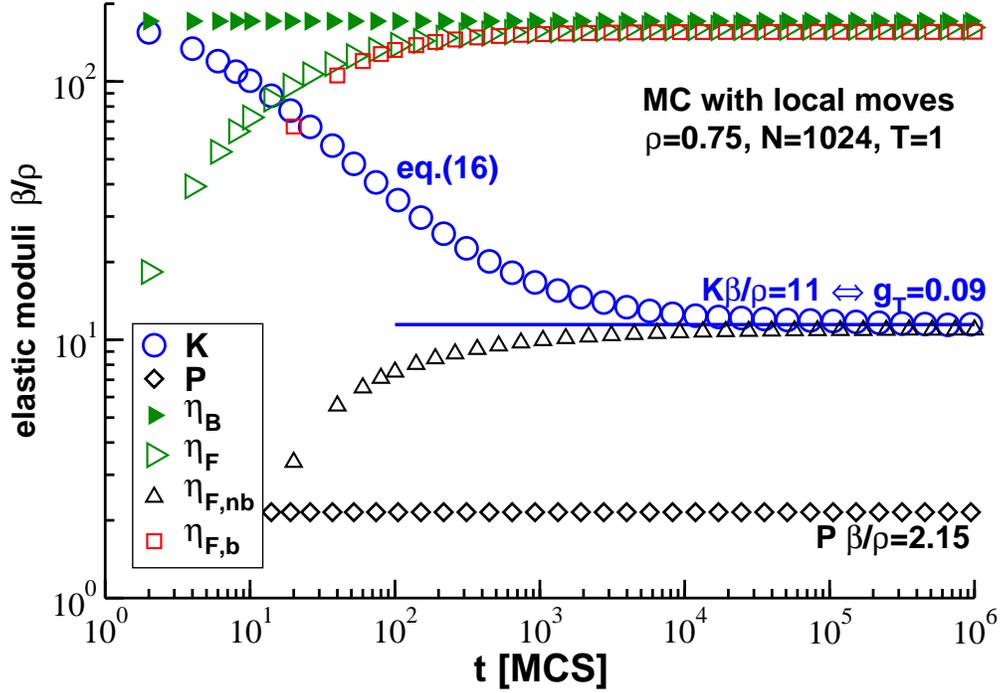}}}
\caption{Elastic properties for $\rho=0.75$ and $N=1024$ as functions of the sampling time $t$:
Compression modulus $K$, total pressure $P$, hypervirial $\Xvirial$, total excess pressure fluctuation $\Xfluctu$
and the contributions $\Xfluctunb$ and $\Xfluctubond$ due to the non-bonded and bonded interactions.
While $P$ and $\Xvirial$ are (essentially) constant, $\Xfluctu(t)$, $\Xfluctunb(t)$ and $\Xfluctubond(t)$ are 
seen to converge only slowly with sampling time and, hence, the compression modulus $K(t)$.
\label{fig_moduli_t}
}
\end{figure}

\begin{figure}[t]
\centerline{\resizebox{.8\columnwidth}{!}{\includegraphics*{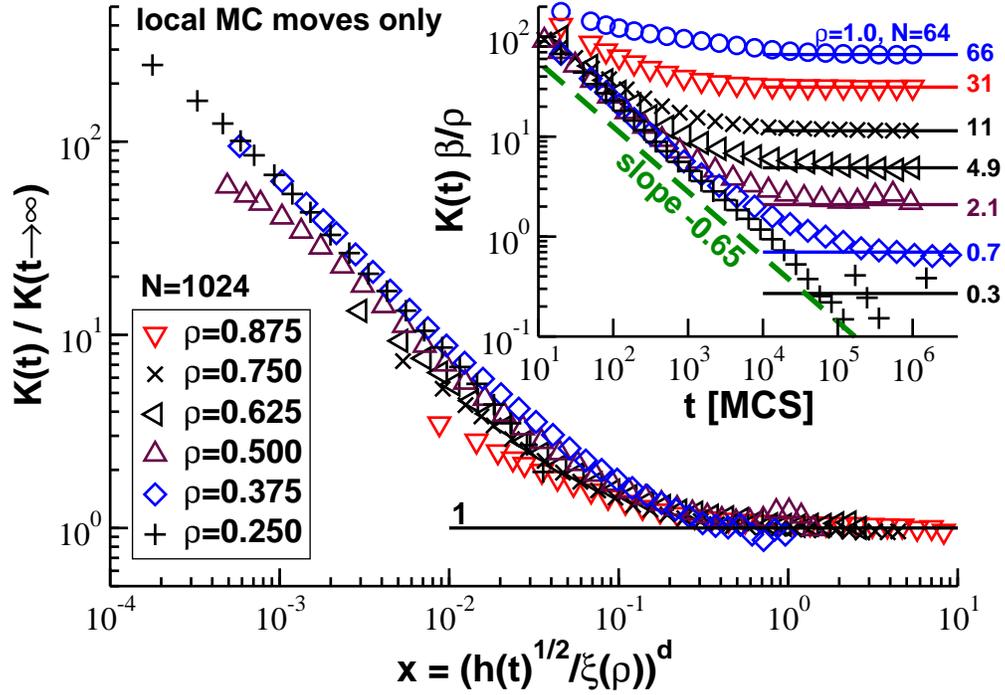}}}
\caption{Monotonous decay of the compression modulus $K(t)$ as a function of the sampling time $t$ 
for different densities $\rho$ for one chain length $N=1024$.
Inset: The smaller the density, the later $K(t) \beta/\rho$ levels off as indicated
by the solid horizontal lines.
Main panel: Scaling plot of $K(t)/K(t\to \infty)$ as a function of the reduced MSD 
$(h(t)^{1/2}/\xi(\rho))^d$
with $\xi(\rho)$ being the blob size assuming eq.~(\ref{eq_blobsize}).
The scaling is not perfect, especially not for larger densities, 
but catches the essential density effect.
\label{fig_Kt_rhoscal}
}
\end{figure}

\paragraph*{Introduction.}
Using different numerical techniques we have determined in the preceding subsection the compression modulus 
$K \beta/\rho = 1/\gTN$ relevant for real experiments. As above for the energy and the pressure
we discuss now in more detail various numerically accessible contributions to $K$.  
One aim is to better analyze the already mentioned numerical difficulties encountered when probing the 
compressibility for densities below $\rhostarstar$ using eq.~(\ref{eq_KRowlinson}). 
Our second aim is to characterize the fluctuations of the (excess) pressure considered in sect.~\ref{res_pressure}.
Various elastic moduli are shown in fig.~\ref{fig_moduli_rho} as functions of the density $\rho$ and 
in fig.~\ref{fig_moduli_t} and fig.~\ref{fig_Kt_rhoscal} as functions of the sampling time $t$. 
All data presented here have been obtained by MC simulation, i.e. there is no discretization 
problem as for MD, and the time is given in MC Steps (MCS) of the local monomer jump attempts \cite{AllenTildesleyBook}.
The vertical axes of the figures are made dimensionless by a factor $\beta/\rho$ 
as in fig.~\ref{fig_energy_rho} and fig.~\ref{fig_pressure_rho},
i.e. we focus on the elastic free energy contribution per monomer.
While we have insisted above on the $N$-dependence due to the translational entropy of the chains, eq.~(\ref{eq_gTNgT}), 
we consider now such high densities and/or large chains that this (additional) 
complication may safely be ignored for the presented data. 

\paragraph*{Density dependence of contributions to $K$.}
The compression modulus $K(\rho)$ obtained using the Rowlinson formula for chains of length $N=1024$ 
replotted in fig.~\ref{fig_moduli_rho} may indeed be considered as $N$-independent. 
Also given are the (rescaled) contributions to $K$ according to eq.~(\ref{eq_KRowlinson}): the mean pressure $P$, 
the hypervirial $\Xvirial$ and the excess pressure fluctuation $\Xfluctu$.
While the pressure is small over the full density range,
$\Xvirial \beta/\rho$ and $\Xfluctu \beta/\rho$ are seen to be 
essentially density independent and of same magnitude.
A reasonable numerical estimation of the compression modulus $K$ thus requires a precise determination of 
its two leading contributions. Note that the logarithmic representation masks the noise naturally present in the data,
especially for lower densities. (This may be better seen from the related data presented in the inset of fig.~\ref{fig_Lame}
given in the Appendix.)
Since with decreasing density $K$ becomes rapidly orders of magnitudes smaller than 
$\Xvirial \approx \Xfluctu$, it is not surprising that we have not been able 
to obtain reliable values for $K$ from eq.~(\ref{eq_KRowlinson}) below $\rho \approx 0.25$.

\paragraph*{Moduli associated to different interactions.}
Applying eq.~(\ref{eq_Xvirial}) and eq.~(\ref{eq_Xfluctu}) to the non-bonded and bonded potential contributions
one obtains the hypervirials $\Xvirialnb$ and $\Xvirialbond$ and the stress fluctuations $\Xfluctunb$ and $\Xfluctubond$. 
Since the hypervirial is linear with respect to the different interactions we have $\Xvirial = \Xvirialnb + \Xvirialbond$,
while $\Xfluctu = \Xfluctunb + \Xfluctubond + \Xfluctumix$ where the last term $\Xfluctumix$ characterizes the correlations between 
the bonded and non-bonded stresses. (Being always much smaller than the other contributions $\Xfluctumix$ is not discussed here.)
As can be seen from the figure we have for all densities
\begin{equation}
\Xvirial \approx \Xfluctu \approx \Xvirialbond \approx \Xfluctubond \gg \Xvirialnb \approx \Xfluctunb,
\label{eq_XbondXnb}
\end{equation}
i.e. the bonded contributions to $\Xvirial$ and $\Xfluctu$ dominate numerically by far the non-bonded interactions.
(Obviously, this does not mean that the difference $\Xvirialnb-\Xfluctunb$ is irrelevant compared
to the difference $\Xvirialbond-\Xfluctubond$.)
The plateau value indicated by the thin horizontal line is thus readily computed from the hypervirial
\begin{equation}
\Xvirialbond \beta/\rho 
= \frac{\beta \kbond}{4} \la r ( 2 r - \lbond ) \ra
\approx \frac{1}{4} \beta \kbond \lbond^2
\label{eq_Xvirialbond}
\end{equation}
with $r$ being the distance between bonded monomers. 
In the second step of eq.~(\ref{eq_Xvirialbond}) it was used that 
$l^2 = \la r^2 \ra \approx \la r \ra \lbond \approx \lbond^2$ 
due to the stiffness of the bonding potential.

\paragraph*{Two-point stress correlations.}
Why is a similar value expected for the fluctuation contribution $\Xfluctu \approx \Xfluctubond$?
To see this we remind first that the fluctuation term, eq.~(\ref{eq_Xfluctu}), may be rewritten
quite generally as
\begin{eqnarray}
\Xfluctu 
& = & \frac{\beta}{d^2V} \sum_{l} \left( \la w(\rijl)^2 \ra - \la w(\rijl)\ra^2 \right) 
\label{eq_Xfluctuselfdef} \\
& + & \frac{\beta}{d^2V} \sum_{l \ne l^{\prime}} \left( \la w(\rijl)w(\rijlprime) \ra - \la w(\rijl)\ra \la w(\rijlprime)\ra \right),
\nonumber
\end{eqnarray}
i.e. the fluctuation contribution to $K$ contains not only two-point correlations ($l=l^{\prime}$)
but also three- and four-point correlations ($l\ne l^{\prime}$).
If we assume that to leading order only the two-point or ``self" correlations matter for
the bonded interactions it follows that
\begin{eqnarray}
\Xfluctubond \beta/\rho & \approx &  
\frac{1}{d^2} \left(\la \beta w(r)^2 \ra - \la \beta w(r) \ra^2 \right)
\label{eq_Xfluctubond_self} \\
& \approx & \left( \frac{\beta \kbond}{d} \right)^2 \la (r-\lbond)^2 r^2 \ra
\label{eq_Xfluctubond_neglect}\\
& \approx & \frac{1}{4} \beta \kbond \lbond^2
\label{eq_Xfluctubond}
\end{eqnarray}
where $r$ stands again for the length of a bond.
In the second step the small squared excess pressure term $\la \beta w(r) \ra^2$ is neglected 
and we have finally taken (again) advantage of the stiffness of the harmonic potential.
The self- or two-point correlation contribution to $\Xfluctu$ can of course be computed directly.
For the total contribution $\Xfluctuself$ and the contribution $\Xfluctuselfbond$ of the bonded interactions
we obtain essentially eq.~(\ref{eq_Xfluctubond}) for all densities. This is not represented
in fig.~\ref{fig_moduli_rho} since these values could not be distinguished (in the double logarithmic 
representation chosen) from the values $\Xvirial$, $\Xfluctu$ already given.
Instead we show the self-contribution $\Xfluctuselfnb$ associated to the non-bonded potential (dashed line)
which is seen to be essentially identical to $\Xvirialnb \approx \Xfluctunb$ \cite{foot_distinct}.

\paragraph*{Time dependence of the compression modulus.}
We have seen above that one difficulty to determine the compression modulus for dilute and semidilute
polymer solutions using the Rowlinson stress fluctuation formula 
stems from the fact that a large hypervirial $\Xvirial$ is essentially compensated by an equally 
large stress fluctuation term $\Xfluctu$. This requires a high precision for determining both contributions.
The computational difficulty is in fact not $\Xvirial$ which is readily obtained
to high precision, but the fluctuation contribution $\Xfluctu$ which is found to require a substantial
subvolume of the configuration space to be sampled.
This point is corroborated in fig.~\ref{fig_moduli_t} presenting various (ultimately static) elastic
contributions as functions of the sampling time $t$ for density $\rho=0.75$.
The presented data have been obtained exclusively using MC simulations with local monomer displacements.
Using time series where instantaneous properties relevant for the moments are written down
every $10$ MCS. All reported properties have been averaged using standard gliding averages 
\cite{AllenTildesleyBook}, i.e. we compute mean values and fluctuations for a given time interval
$[t_0,t_1=t_0+t]$ and average over all possible intervals of length $t$.
It is seen that simple means such as the pressure $P$ and the hypervirial $\Xvirial$ reach immediately 
their asymptotic values.
The open circles correspond to $K$ computed according to eq.~(\ref{eq_KRowlinson}) for a canonical ensemble 
at constant volume. 
Measuring fluctuations $\Xfluctu$ must vanish if only one configuration is measured ($t=0$).
Since $\Xfluctu(t)$ {\em increases} for small times, $K$ approaches the asymptotic value
from above requiring about $10^5$ MCS to reach the plateau.
(That the fluctuation contribution $\Xfluctunb(t)$ of the non-bonded interactions becomes similar for large times
is by accident for the density presented in the figure.)

\paragraph*{Density scaling of time dependence.}
The time dependence of $K(t)$ is further investigated in fig.~\ref{fig_Kt_rhoscal} 
where we compare different densities $\rho$ for $N=1024$.
As shown in the inset $K(t) \beta/\rho$ decreases monotonously both with decreasing density and increasing sampling time.
(We have included here in addition a data set for $\rho=1.0$ and $N=64$.)
The plateau values $K(t\to \infty)$ for long times are marked by the horizontal lines. It is seen that increasingly
more time is needed to reach the plateau if the density is lowered \cite{foot_isobaric}. 

The $\rho$-scaling of the crossover time is clarified in the main panel where we have plotted the rescaled vertical axis 
$K(t)/K(t\to \infty)$ as a function of the measured mean-square displacement (MSD) $h(t)$ for each density.
Hence, instead of the sampling time we use the volume $h^{d/2}(t)$ explored by the beads as coordinate.
The horizontal axis is then made dimensionless by reducing the MSD using the blob size $\xi(\rho)$ given by eq.~(\ref{eq_blobsize}).
As can be seen we observe a satisfactory scaling collapse for all data sets. 
We emphasize that the time-independent thermodynamic limit 
is reached when the beads have explored the semidilute blob, i.e. $h(t) \approx \xi^2(\rho)$.
That the collapse is not perfect for our highest densities $\rho > 0.5$ is expected since the
semidilute blob scaling breaks down in this limit. 
Data for lower densities are warranted in the future to demonstrate the suggested blob scaling unambiguously.

\begin{figure}[t]
\centerline{\resizebox{0.8\columnwidth}{!}{\includegraphics*{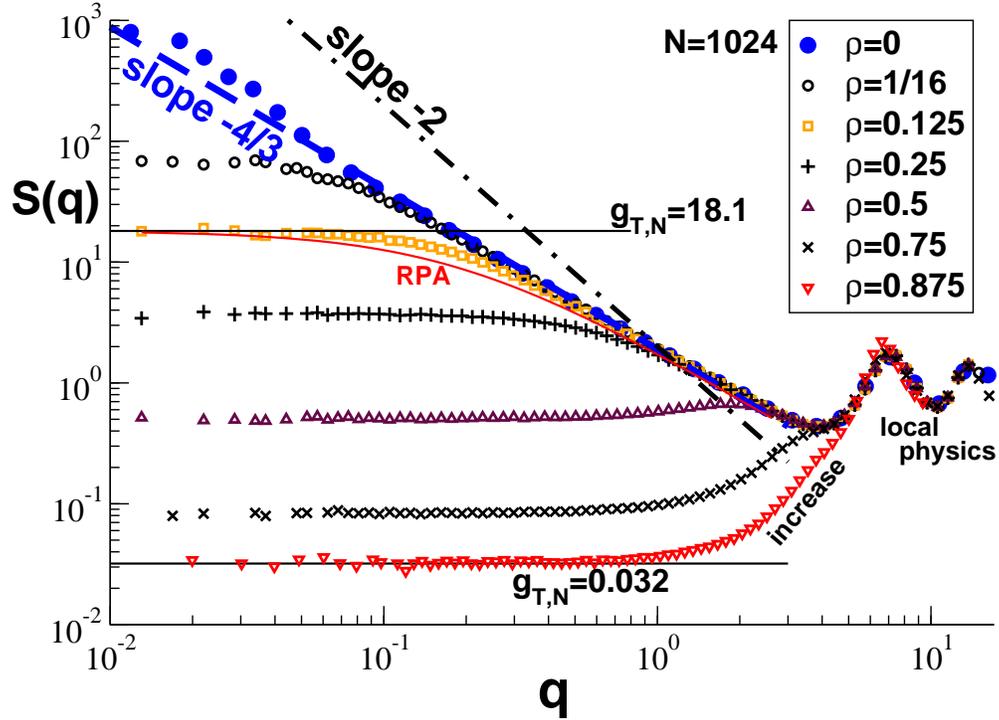}}}
\caption{Total monomer structure factor $S(q)$ {\em vs.} wavevector $q$ for $N=1024$ and several densities $\rho$.
In the dilute limit $S(q)$ corresponds to the single chain form factor $\Fdil(q)$ 
indicated by the small filled circles. The plateau values in the low-$q$ limit for finite $\rho$
have been used for the determination of $\gTN(\rho)$ as shown for $\rho=0.125$ and $\rho=0.5$ (thin horizonal lines).
At smaller wavevectors $S(q)$ is found to decay monotonously only for densities $\rho < 0.5$. 
At higher densities $S(q)$ increases with $q$ in qualitative difference to the RPA formula, eq.~(\ref{eq_Sq_RPA}), 
which is indicated by the thin line for $\rho=0.125$ using the directly measured intramolecular structure factor $F(q)$. 
\label{fig_Sq_rho}
}
\end{figure}

\subsection{Total monomer structure factor $S(q)$}
\label{res_Sq}
The compression modulus $K$ discussed above describes
the linear response of the systems in the hydrodynamic limit for wavevectors $q \to 0$.
We turn now to an experimentally highly relevant reciprocal space characterization
of the polymer solution characterizing the compressibility at a given wavevector $q$:
the ``total monomer structure factor" \cite{DegennesBook,mai00}
\begin{equation}
S(q) = \frac{1}{\nmon} \sum_{n,m=1}^{\nmon}
\langle \exp\left[i \qvec \cdot (\rvec_n-\rvec_m) \right]\rangle. 
\end{equation}
In our simulations all $\nmon$ monomers of the simulation box are assumed to be labeled and the average
$\la \ldots \ra$ is performed over all configurations of the ensemble and all possible 
wavevectors of length $q = |\qvec|$. 
As already discussed in sect.~\ref{res_compress}, the isothermal compressibility of the solution may be 
obtained from the plateau of the structure factor in the low-$q$ limit according to eq.~(\ref{eq_Sq2gT}). 
The smallest possible wavevector is of course $2\pi/\Lbox$ since the wavevector $\qvec$ must be commensurate 
with the simulation box. As can be seen from fig.~\ref{fig_Sq_rho}
our box sizes allow for a precise determination of $\gTN(\rho)$ for all densities $\rho \ge 0.0625$.
Only chains of length $N=1024$ are presented for clarity.
Since above $q > 3$ the monomer structure becomes important for all densities
and, being interested in universal physics, we focus below on smaller wavevectors.
For comparison we have also included the intramolecular structure factor $F(q)=\Fdil(q)$ 
obtained for one single chain in a large box ($\rho = 0$). As emphasized by the dashed line
$\Fdil(q)$ is characterized in the intermediate wavevector regime by the power-law decay 
\begin{equation}
\Fdil(q) = \frac{\cdil}{(\bdil q)^{4/3}} 
\mbox{ for } \frac{1}{\Rgyr} \ll q \ll \frac{1}{\sigma}
\label{eq_Fqdil}
\end{equation}
with $\cdil \approx 0.5$ being a dimensionless amplitude and $\sigma$ representing the monomer scale.

Returning to finite monomer densities we remind the ``random phase approximation" (RPA) \cite{DegennesBook}
\begin{equation}
\frac{1}{S(q)} = \frac{1}{\gT} + \frac{1}{F(q)}
\label{eq_Sq_RPA}
\end{equation}
with $F(q)$ being the intramolecular structure factor at the given density.
The RPA is supposed to relate --- at least for not too high densities ---
the total structure factor $S(q)$ to the dilute chain form factor $F(q) \approx \Fdil(q)$
for $q \gg 1/\xi$ \cite{DegennesBook,WCX11}.
Note that $\gT$ stands for the excess contribution to the dimensionless compressibility 
in agreement to eq.~(\ref{eq_gTNgT}) and to $1/F(q) \to 1/N$ in the low-$q$ limit.
By construction $S(q)$ is fitted by the RPA in the low-$q$ and large-$q$ limits
as shown by the thin solid line for $\rho=0.125$. Interestingly, even for low densities the 
crossover regime between both $q$-limits at $q \approx 2\pi/\xi$ is, however, only inaccurately described
\cite{foot_RPAbulk}.
The monotonous decay of $S(q)$ implicit to eq.~(\ref{eq_Sq_RPA}) is indeed observed for
the semidilute densities with $\rho < 0.5$. For higher densities, where the blob picture
breaks down due to monomer physics, $S(q)$ becomes essentially constant for all $q$ up to the 
monomeric scale.
\begin{figure}[t]
\centerline{\resizebox{0.8\columnwidth}{!}{\includegraphics*{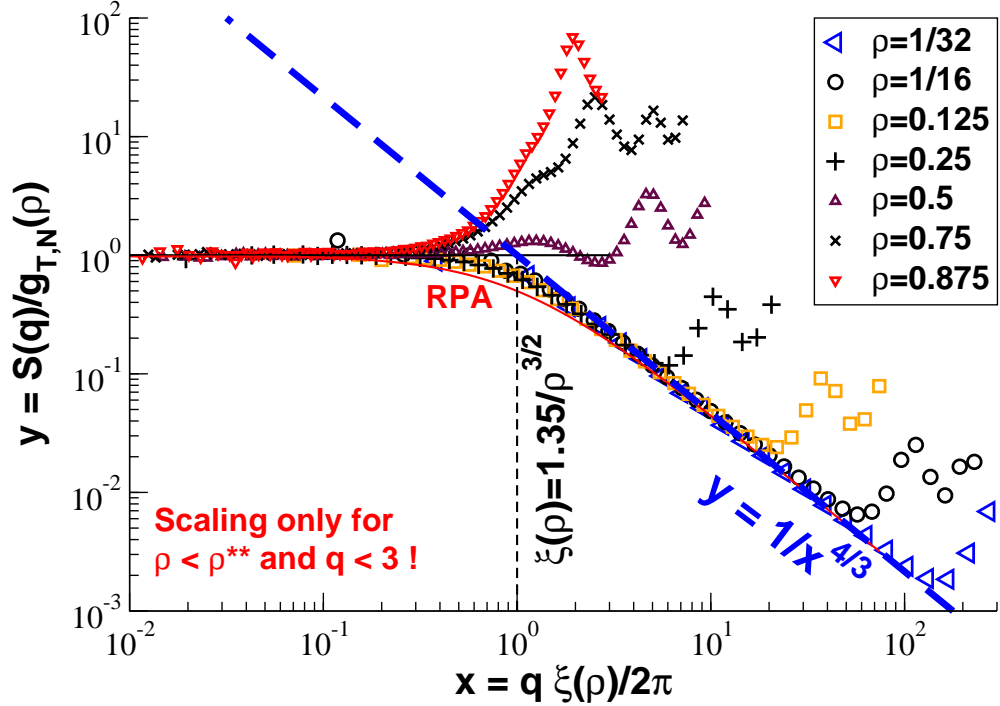}}}
\caption{Rescaled structure factor $y = S(q)/\gTN(\rho)$ as a function of reduced wavevector $x = q \xi(\rho)/2\pi$.
Assuming eq.~(\ref{eq_blobsize}) for the blob size $\xi$, the asymptotic small-$q$ (horizontal line) and 
large-$q$ (bold dashed line) limits match at $x \approx 1$ as indicated by the vertical dashed line.
\label{fig_Sq_scal}
}
\end{figure}

The scaling of $S(q)$ in the semidilute regime is further investigated in fig.~\ref{fig_Sq_scal}
where  $y = S(q)/\gTN$ is plotted as a function of $x = q \xi/2\pi$ with $\xi$ being set 
by the matching of the radius of gyration in the dilute and semidilute limits, eq.~(\ref{eq_blobsize}).
As indicated by the vertical dashed line the asymptotic slopes for small and large wavevectors
are found to intercept at $x = 1$ ! Using eq.~(\ref{eq_Fqdil}) for the dilute chain form factor
this implies $\xi = \bdil 2\pi (\gT/\cdil)^{\nudil}$. Since on the other side $\xi = \bdil g^{\nudil}$,
it follows that the number $g$ of monomers spanning the blob is given by $g \approx 23 \gT$ 
as already stated above, eq.~(\ref{eq_gTsemidilute}).
Our operational prefactor setting, eq.~(\ref{eq_blobsize}), thus corresponds to an experimentally 
(in principle) measurable choice. It suggests that future experimental work may proceed in a similar 
manner by fixing $\xi$ (or equivalently $g$) from the matching point of both $q$-limits of the total
structure factor rather than by  imposing an inappropriate Ornstein-Zernike
fit to the intramolecular form factor $F(q)$ or the total structure factor $S(q)$ \cite{mai00}.

\section{Conclusion}
\label{sec_conc}

{\em Summary.}
In this paper we investigated numerically the crossover scaling of various thermodynamic properties 
of solutions and melts of a generic bead-spring model of self-avoiding and highly flexible polymer 
chains without chain intersections confined to strictly two dimensions.
As expected \cite{Duplantier89,ANS03,SMW12} the typical interaction energy $\einter$ between monomers 
from different chains was shown (fig.~\ref{fig_energy_N}) to scale as $\einter \sim 1/N^{\nu \thetatwo}$ 
with $\nu=3/4$ and $\thetatwo=19/12$ for dilute solutions 
and $\nu=1/d=1/2$ and $\thetatwo=3/4$ for sufficiently large chains with $N \gg g(\rho) \sim 1/\rho^2$ 
where the chains adopt compact configurations of fractal perimeter dimension.
In the semidilute regime we confirmed over an order of magnitude in density $\rho$
the power-law scaling 
for the interaction energy $\einter \sim \rho^{21/8}$ (fig.~\ref{fig_energy_rho}),
the pressure $P \sim \rho^3$ (fig.~\ref{fig_pressure_scal})
and the dimensionless isothermal compressibility $\gT \sim 1/\rho^2$ (fig.~\ref{fig_compress})
expected theoretically in the limit of asymptotically long chains.
Polymer specific technical difficulties associated with the numerical determination of pressure
(inset of fig.~\ref{fig_pressure_scal}) and elastic moduli 
(figs.~\ref{fig_moduli_rho}-\ref{fig_Kt_rhoscal} and \ref{fig_Lame}) 
at small and intermediate densities have been discussed.
The elastic contributions $\Xvirial$ and $\Xfluctu$ associated, respectively, to the affine and non-affine response 
to an imposed homogeneous strain were analyzed as functions of density and sampling time. We have emphasized that 
\begin{equation}
\Xvirial \approx \Xfluctu \approx \Xvirialbond \approx \Xfluctubond 
\approx \kbond \lbond^2 /4\rho \gg K \gg P
\label{eq_diffproblem}
\end{equation}
for all but the highest densities \cite{foot_distinct}.
(Similar relations exist for the Lam\'e coefficients are discussed in Appendix~\ref{app_Lame}.) 
The stress fluctuations $\Xfluctu(t)$ become only time independent if distances corresponding to the 
blob size $\xi(\rho)$ are probed (fig.~\ref{fig_Kt_rhoscal}).
Returning finally to experimentally relevant properties we showed how the size of the semidilute blob $\xi(\rho)$ 
may be determined in a real experiment from the pressure isotherms, eq.~(\ref{eq_P2xi}), or
the total monomer structure factor $S(q)$ characterizing the compressibility 
of the solution at a given wavevector $q$ (fig.~\ref{fig_Sq_scal}).
\begin{figure}[t]
\centerline{\resizebox{.8\columnwidth}{!}{\includegraphics*{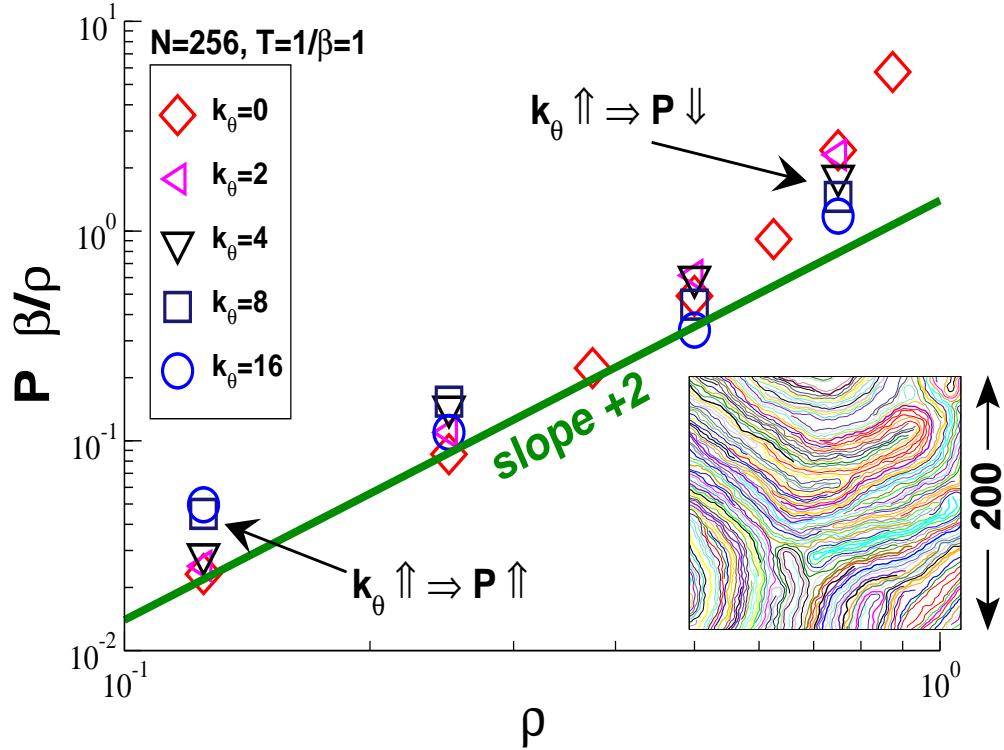}}}
\caption{Reduced pressure $P \beta/\rho$ {\em vs.} density $\rho$ for $N=256$ and several stiffness energy penalties $\kstiff$.
In agreement with previous computational findings \cite{dijkstra94} the pressure $P(\kstiff)$ increases with $\kstiff$  
for low densities but {\em decreases} for high densities.
Inset: Snapshot of a subvolume of a larger box showing local order and hairpin-like defects
obtained for $N=256$, $\rho=0.5$, $\Lbox \approx 313$ and $\kstiff=16$ starting from an isotropic
and flexible system with $\kstiff=0$. Due to the hairpins the equilibration dynamics
becomes sluggish and strong hysteresis effects are observed if systems with different
dynamical pathways are compared. 
\label{fig_pressure_stiff}
}
\end{figure}

{\em Polymer blends.}
As already argued elsewhere \cite{ANS03,SMW12}, the presented numerical results for the interchain interaction energy 
$\einter$ for homopolymer systems implies an enhanced compatibility for polymer blends confined to ultrathin films
in qualitative agreement with recent experimental studies \cite{Kumaki12}.
We remind that the critical temperature of unmixing $\Tcrit$ should be proportional to the typical 
interaction energy $N \einter$ between different chains. According to eq.~(\ref{eq_ninter_rho}) and 
using eq.~(\ref{eq_grho}) it follows thus that $\Tcrit \sim \rho^{21/8} N^{5/8}$ in the compact chain limit.
This scaling is in fact consistent with recent MC simulations of symmetrical polymer
mixtures using a version of the bond-fluctuation model in strictly two dimensions \cite{Cavallo03,Cavallo05}.
The predicted exponent $5/8=0.625$ for the $N$-dependence agrees well with the value $0.65$ obtained by fitting 
$\Tcrit$ for all computed chain lengths \cite{Cavallo03}.
The corresponding strong power-law increase $\Tcrit \sim \rho^{21/8}$ with density 
has to our knowledge not been probed yet which calls for an experimental and/or numerical
verification focusing on semidilute polymer blends.

{\em Persistence length effects.}
Returning to homopolymer solutions we note finally that, at variance to most experimental systems 
\cite{mai00,Rubinstein07,Monroy11},
we have assumed here that the chains are flexible down to monomeric scales. As long as the persistence length remains smaller 
than the blob size, this should in fact not alter the suggested scaling properties.
In order to bring our computational approach closer to experiment we are currently computing systems with finite persistence length.
In addition to the coarse-grained model Hamiltonian presented in sect.~\ref{sec_algo} a local stiffness potential 
$\kstiff ( 1 - \cos(\theta))$ is applied with $\theta$ being the angle between adjacent bonds of a chain.
As shown by the pressure isotherms presented in fig.~\ref{fig_pressure_stiff} for chains of length $N=256$
the scaling with respect to density for sufficiently long chains remains essentially unchanged.
In fact, since the blob size decreases very strongly with persistence length, i.e. $N/g$ increases, 
a finite rigidity even speeds up the convergence to the predicted asymptotic behavior.
As expected from the decreasing blob size $\xi$, the pressure $\beta P \approx 1/\xi^d$
is seen to {\em increase} with $\kstiff$ for small densities.
Obviously, if the stiffness penalty $\kstiff$ and/or the density $\rho$ become too large,
a nematic chain alignment becomes relevant (at least locally) as shown by the snapshot shown in fig.~\ref{fig_pressure_stiff}.
Due to this additional effect $P$ is found to {\em decrease} with $\kstiff$ for large $\rho$ in agreement with the numerical 
study by Dijkstra and Frenkel \cite{dijkstra94}.

\begin{acknowledgments}
A grant of computer time by the ``P\^ole Mat\'eriaux et Nanosciences d'Alsace" (ENIAC) is gratefully acknowledged.
N.S. thanks the R\'egion d'Alsace for financial support, 
P.P. the IRTG Soft Matter,
H.X. the CNRS and the IRTG Soft Matter for supporting her sabbathical stay in Strasbourg.
We are indebted to C. Marques and T. Charitat (all ICS, Strasbourg) for helpful discussions.
\end{acknowledgments}

\appendix

\begin{figure}[t]
\centerline{\resizebox{.8\columnwidth}{!}{\includegraphics*{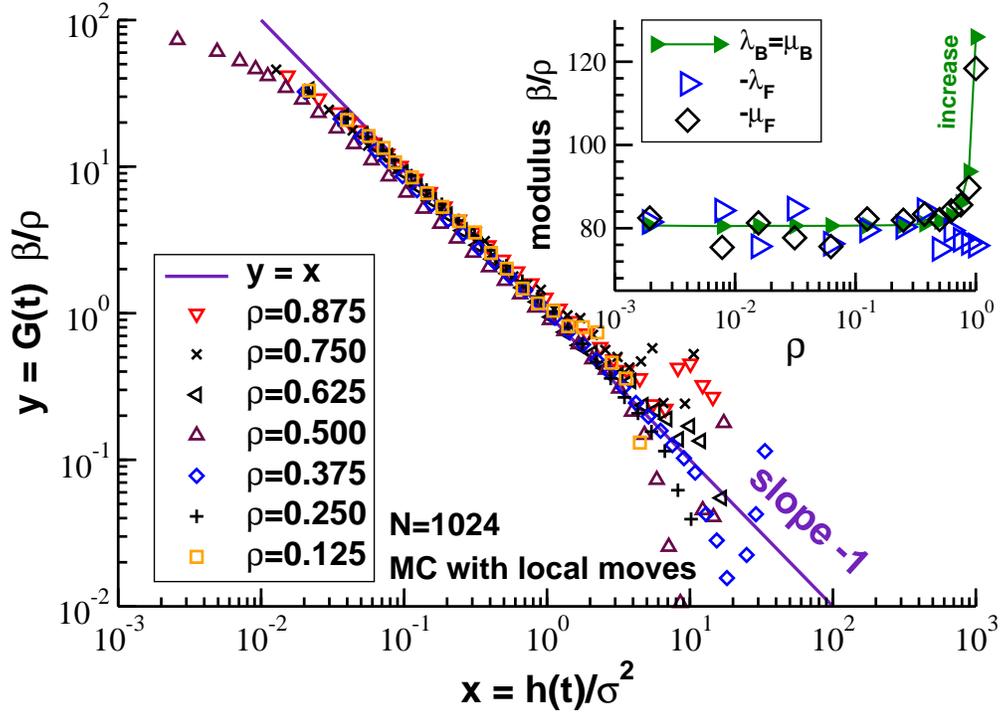}}}
\caption{Lam\'e coefficients and related properties for $N=1024$. 
The vertical axes is made dimensionless by means of a factor $\beta/\rho$.
Inset: Born Lam\'e coefficients $\laB=\muB$ and stress fluctuations $-\laF$ and $-\muF$ as functions of $\rho$. 
The data for dilute and semidilute densities is well approximated by eq.~(\ref{eq_laBlaFmuFplateau}).
The vertical axis being linear, the noise appears more strongly than for $\Xvirial$
and $\Xfluctu$ in fig.~\ref{fig_moduli_rho} where a logarithmic representation is chosen.
(The data point for $\rho=1.0$ has been obtained for $N=64$.)
Main panel: Time dependent shear modulus $G(t) = \muB + \muF(t) - \Pex$ for several densities 
as a function of MSD $h(t)$. The data for different densities scale if a density
independent local scale $\sigma$ is assumed for the horizontal axis.
\label{fig_Lame}
}
\end{figure}

\section{Lam\'e coefficients and related properties}
\label{app_Lame}

\paragraph*{Definitions.}
For consistency with previous numerical work \cite{WTBL02,SBM11} we present here the different contributions to
the Lam\'e coefficients $\lambda$ and $\mu$ of the solution \cite{LandauElasticity}. As may be seen by 
thermodynamic and symmetry considerations, the compression modulus $K$ and the shear modulus $G$ 
of any isotropic and homogeneous system in $d$ dimensions may be rewritten as 
\begin{eqnarray}
K & = & (\lambda + P) + \frac{2}{d} G, \label{eq_lame2K}\\
G & = & \mu - P \label{eq_lame2G}
\end{eqnarray}
where we follow the notation of ref.~\cite{SBM11}. 
We do this to emphasize the explicit
pressure dependence which is often (incorrectly) omitted \cite{LandauElasticity} 
as clearly pointed out by Birch \cite{Birch37} and Wallace \cite{Wallace70}.
Note that $K = \lambda + \mu$ in $d=2$ dimensions. 
Since by definition of a liquid the shear modulus must vanish for our systems, $G = 0$, 
the Lam\'e coefficient $\mu$ is simply given by the total pressure $P$
(as indicated in fig.~\ref{fig_moduli_rho}). Hence,
\begin{eqnarray}
K & = & \Pid + \Pex + \laB - (-\laF) \label{eq_KlaBlaF} \\
0 = G & = & \muB - (-\muF) - \Pex \label{eq_muBmuFPex} 
\end{eqnarray}
where we have rewritten the Lam\'e coefficients as \cite{SBM11}
\begin{eqnarray}
\lambda & \equiv & \laB - (-\laF)     \nonumber,  \\
\mu - \Pid  & \equiv & \muB - (-\muF) \label{eq_lamu2BF}.
\end{eqnarray}
Please note that the only contribution due to the kinetic energy of the particles
is contained by the ideal gas pressure $\Pid$ indicated for $\mu$. 
Kinetic energy contributions to the elastic moduli are removed as far as possible
in view of the fact that MC results are considered here.

\paragraph*{Born Lam\'e coefficients.}
The first contributions indicated on the right hand-side of eq.~(\ref{eq_lamu2BF}) 
are the so-called ``Born Lam\'e coefficients"
%
\begin{equation}
\laB \equiv \muB  
\equiv \frac{1}{V} \la \sum_{l} \left( \rijl^2 u^{\prime\prime}(\rijl) - \rijl u^{\prime}(\rijl) \right) \frac{\xijl^2 \yijl^2}{\rijl^4} \ra  
\label{eq_lameBorn}
\end{equation}
where the index $l$ stands for the interaction between two monomers $i < j$,
$\rijl$ for the length of the vector $\rijlvec$ between both monomers
and $\xijl$ and $\yijl$ for its components.
The Born Lam\'e coefficients characterize the free energy change of the systems assuming an {\em affine} displacement 
of all particles due an imposed external homogeneous linear strain \cite{Lutsko89,WTBL02}. 
Using symmetry considerations it can be readily seen that
\begin{equation}
\Xvirial = \laB + \frac{2}{d} \left( \muB - \Pex \right),
\label{eq_laB2Xvirial}
\end{equation}
i.e. $\laB = (\Xvirial + \Pex)/2 \approx \Xvirial/2$ as confirmed by the data indicated in Table~\ref{tab_rho}.
As shown in the inset of fig.~\ref{fig_Lame}, the Born Lam\'e coefficients per monomer (filled triangles)
are constant for dilute and semidilute densities, but increase strongly for our largest melt densities.

\paragraph*{Stress fluctuations.}
The two remaining terms $\laF$ and $\muF$ in eq.~(\ref{eq_lamu2BF}) characterize the 
fluctuations of the stress tensor
\begin{eqnarray}
-\laF & \equiv & \beta V \la \delta \hat{P}_{xx} \delta \hat{P}_{yy} \ra  \label{eq_laF}\\
-\muF & \equiv & \beta V \la \delta \hat{P}_{xy} \delta \hat{P}_{xy} \ra \label{eq_muF}
\end{eqnarray}
with $\hat{P}_{\alpha\beta}$ being the (instantaneous) excess pressure tensor
and $\delta \hat{P}_{\alpha\beta} \equiv \hat{P}_{\alpha\beta} - \la \hat{P}_{\alpha\beta} \ra$ 
a fluctuation.
We remind that the mean excess pressure $\Pex$ is the averaged trace over the 
instantaneous excess pressure tensor, $\Pex = \la \Trace[ \hat{P}_{\alpha\beta}] \ra/d$.
The above definitions do also hold in higher dimensions $d > 2$ albeit averages may be performed there
over equivalent pairs of spatial indices. The minus sign in front of $\laF$ and $\muF$ is introduced
to be consistent with related work \cite{WTBL02,SBM11,paptrunc}.
By writing the virial $\Wvirial/V$ in the definition of $\Xfluctu$, eq.~(\ref{eq_Xfluctu}),
in terms of the diagonal elements $\delta \hat{P}_{\alpha\alpha}$ of the pressure tensor  
one sees using symmetry considerations that
\begin{equation}
\Xfluctu = (-\laF) + \frac{2}{d} (-\muF).
\label{eq_laF2W}
\end{equation}
The reader may verify that eq.~(\ref{eq_laB2Xvirial}) and eq.~(\ref{eq_laF2W}) 
are consistent with eq.~(\ref{eq_KRowlinson}) and eq.~(\ref{eq_lame2K}).
Note that eq.~(\ref{eq_muBmuFPex}) implies $\muB \approx -\muF$ for low densities where $\Pex$ is negligible. 
While $\laB \equiv \muB$ by definition for pairwise central forces the Lam\'e coefficients $\laF$ and $\muF$ may differ in general.
As can be seen from the inset of fig.~\ref{fig_Lame} (open symbols) or Table~\ref{tab_rho} we obtain, however, to leading order
\begin{eqnarray}
\laB \beta/\rho & \approx & - \laF \beta/\rho \approx -\muF \beta/\rho \label{eq_laBlaFmuF} \\
 & \approx & \beta \kbond \lbond^2/8 \approx 80
\label{eq_laBlaFmuFplateau}
\end{eqnarray}
for $\rho \le 0.5$. 
In the last step we have used the same reasoning as in eq.~(\ref{eq_Xvirialbond}) assuming that
the bonding potential dominates the moduli for low densities.
That eq.~(\ref{eq_laBlaFmuF}) holds (albeit not rigorously)
can again be traced back to the fact that strong two-point (self) interaction contributions
to $\Xfluctu$ (especially due the bonding potential) dominate over the distinct correlations 
$\Xfluctudist$ for distinct interactions ($l \ne l^{\prime}$).
Using eq.~(\ref{eq_laF2W}) this implies to a good approximation $-\laF \approx -\muF \approx \Xfluctu/2$.
Coming back to eq.~(\ref{eq_KlaBlaF}) one sees that $K$ is again represented by the difference of two large terms $\laB$ and $-\laF$
of same magnitude. Considering the statistics of the data visible in the inset of figure, it is clear that our $\laF$-values 
do not allow to extend the computation of $K$ to smaller densities, just as is was the case using the Rowlinson formula.

\paragraph*{Average time-dependent shear modulus.}
As the hypervirial $\Xvirial$ the Born Lam\'e coefficients $\laB=\muB$ reach immediately their asymptotic values (not shown). 
This is again different for the stress fluctuation contribution $\muF$ as shown in the main panel of fig.~\ref{fig_Lame}
which presents the time dependent shear modulus $G(t)$. We use here that according to eq.~(\ref{eq_lame2G}) and 
eq.~(\ref{eq_lamu2BF}) the shear modulus is given by \cite{SBM11} 
\begin{equation}
G(t) \equiv \muB - (-\muF(t)) - \Pex.
\label{eq_Gt}
\end{equation}
As it should for a liquid, $G(t)$ vanishes rapidly \cite{foot_truncated,paptrunc}. 
The vertical axis is made dimensionless by tracing the modulus per monomer $G(t) \beta/\rho$. 
As horizontal axis we use the monomer MSD $h(t)$.
Interestingly, the data collapse for different densities although we have {\em not} rescaled the axis 
with the blob size $\xi(\rho)$ as we did for the compression modulus $K(t)$ in fig.~\ref{fig_Kt_rhoscal}. 
Density effects enter here only through the density dependence of
the monomer mobility which we have scaled out by using the measured MSD $h(t)$ as coordinate.
That the scaling of $K(t)$ depends {\em explicitly} on the density 
--- and thus according to eq.~(\ref{eq_KlaBlaF}) the Lam\'e coefficient $\laF(t)$ ---
stems from the fact that the compression modulus couples to the density fluctuations 
which are characterized by $\xi(\rho)$.
Not coupling directly to the density fluctuations, the shear modulus $G(t)$ and thus $\muF(t)$ 
decay on a local monomer scale. 
%
%

\clearpage

\begin{thebibliography}{54}
\bibitem[{foo({\natexlab{a}})}]{foot_approx}
\bibinfo{note}{If we compute a quantity exactly, including all numerical
  coefficients, we use an equals sign, i.e., we write $A=B$. If we state only a
  scaling law, ignoring all numerical coefficients, but keeping all dimensional
  factors, we use the symbol $\approx$ as, e.g., for the chain size in the
  compact limit, Eq.~(\ref{eq_compact}). If we want to stress only the power
  law involved, we use the symbol $\sim$ as, e.g., for the scaling of the chain
  perimeter $L$ with chain length $N$ indicated in Eq.~(\ref{eq_dperi}). The
  dilute limit of a property considered is often characterized by an index $0$,
  e.g., $\nudil=3/4$ denotes the Flory chain size exponent for dilute chains.}
\bibitem[{foo({\natexlab{b}})}]{foot_truncated}
\bibinfo{note}{Being truncated and shifted at the minimum of the full LJ
  potential our excluded volume potential is continuous and differentiable
  everywhere. As shown in Ref.~\cite{paptrunc}, this is of relevance for
  calculations of elastic moduli using a stress fluctuation relation, such as
  eq.~(\ref{eq_KRowlinson}), which involves derivatives of the interaction
  potentials.}
\bibitem[{foo({\natexlab{c}})}]{foot_connectedmons}
\bibinfo{note}{This clearly separates the bonded and non-bonded interactions
  which is of importance for the various thermodynamic contributions
  investigated in Sect.~\ref{sec_res}. Note that some implementations of the KG
  model, as the recent version of the LAMMPS code, allow to view the LJ
  interactions between bonded monomers as intrachain contributions.}
\bibitem[{foo({\natexlab{d}})}]{foot_harmonic}
\bibinfo{note}{The bond potential being harmonic, various conformational and
  thermodynamic properties can easily be calculated if the non-bonded potential
  is thought to be switched off or known to be irrelevant. Under this
  assumption the equipartition theorem \cite{AllenTildesleyBook} tells us,
  e.g., that the average bonding energy $\ebond$ per bond should be $\kBT/2$.
  As a consequence the relative deviation from the reference distance $\lbond =
  0.967$ is given by $$\la (r/\lbond - 1)^2 \ra^{1/2} = \sqrt{\kBT/\kbond
  \lbond^2} \approx 0.0398.$$ This gives an excellent approximation for the
  data in the dilute and semidilute regimes where $\la |r| \ra \approx 0.9692$
  and $l \equiv \la r^2 \ra^{1/2} \approx 0.9700$ (Table~\ref{tab_rho}).}
\bibitem[{foo({\natexlab{e}})}]{foot_comppower}
\bibinfo{note}{The reported MD data have been sampled over a period of about
  five years using different local and national computational resources which
  are difficult to compare. The configurations obtained in this limit have been
  already used and characterized in various previous publications
  \cite{MKA09,MWK10,MSZ11,SMW12,WMJ10}. The production of the MC data for
  smaller densities ($\rho \le 0.25$) performed to crosscheck and improve the
  MD simulations was much less expensive corresponding to production runs over
  a year using 8 cores of Intel Xeon E5410 processors.}
\bibitem[{foo({\natexlab{f}})}]{foot_Pphantom}
\bibinfo{note}{For non-interacting phantom chains we have $-\kBT$ for the
  pressure contribution per bond as one confirms by integration by parts of
  $\int \ddiff r r \ubond^{\prime}(r) e^{-\beta \ubond(r)}$. Summing over all
  bonds we thus have $\Pbond \beta/\rho = -(N-1)/N=-1+1/N$ and, hence, $P =
  \Pid + \Pbond = \kBT \rho/N$.}
\bibitem[{foo({\natexlab{h}})}]{foot_Pbest}
\bibinfo{note}{Since the non-bonded interactions get more important at higher
  densities, these numerical problems become irrelevant for $\rho \ge 0.25$.
  The data points given in fig.~\ref{fig_pressure_rho} and the main panel of
  fig.~\ref{fig_pressure_scal} all refer to the best $\delta t$-independent
  thermodynamic relevant values available.}
\bibitem[{foo({\natexlab{i}})}]{foot_Pscal}
\bibinfo{note}{This scaling has been directly tested by tracing $N P
  \beta/\rho$ as a function of $x = \rho/\rhostar \sim \rho N^{1/2}$. This plot
  is not presented since the related dilute-semidilute crossover scaling for
  the compressibility is given in the inset of fig.~\ref{fig_compress}.}
\bibitem[{foo({\natexlab{j}})}]{foot_isobaric}
\bibinfo{note}{We have additionally checked that similar values are obtained
  from the volume fluctuations $\delta V$ in an isobaric ensemble with imposed
  pressure $P$ using $K = \kBT \la V \ra/\la \delta^2 V\ra$
  \cite{AllenTildesleyBook}. While we find again that this method is
  straightforward for polymer melts ($\rho > 0.5$), $K(t)$ is seen to converge
  increasingly slowly with decreasing density to the asymptotic long-time
  plateau --- just as the compression moduli computed using the stress
  fluctuation formula, eq.~(\ref{eq_KRowlinson}), for the canonical ensemble
  presented in fig.~\ref{fig_Kt_rhoscal}.}
\bibitem[{foo({\natexlab{l}})}]{foot_RPAbulk}
\bibinfo{note}{A similar deviation of the RPA formula in the crossover regime
  at $q \approx 2\pi/\xi$ has also been seen for three-dimensional bulks
  \cite{MM00}.}
\end{thebibliography}


\begin{table}[t]
\begin{tabular}{|c||c|c|c|c|c|c|c|c|c|c|c|c|c||c|c|}
\hline
$\rho$                           &
$M$                              & 
$l$                              &
$\Rend$                          &
$\Rgyr$                          &
$\frac{\beta \enb}{10^{-2}}$     &
$\frac{\beta \einter}{10^{-6}}$  &
$P \beta/\rho$                   & 
$\gTN$                           &
$\Xvirial \beta/\rho$            &
$\Xfluctu \beta/\rho$            &
$\laB \beta/\rho$                &
$-\laF \beta/\rho$               &
$-\muF \beta/\rho$               &
$\gT$                            &
$\xi$                            \\ \hline
1/128 &96 &0.970&173&65  &1.6&0.25 &0.001 &$811^{\star}$ &162&167&84.2&80.5&75.4&$3900^{\star}$&$1955^{\star}$   \\
1/64  &48 &0.970&170&65  &1.6&0.96 &0.001 &$500^{\star}$ &162&145&61.2&80.6&81.3&$975^{\star}$&$691^{\star}$    \\
1/32  &96 &0.970&167&64  &1.6&1.9  &0.002 &$197^{\star}$ &162&170&92.6&80.6&77.7&$244^{\star}$&$244^{\star}$     \\
1/16  &48 &0.970&148&58  &1.6&6.4  &0.005 & 67           &162&155&76.3&80.6&75.6& 67          &86                 \\
0.125 &48 &0.969&121&50  &1.8&37   &0.024 & 18           &162&162&85.0&80.6&78.1& 18          &31                 \\
0.250 &96 &0.969&87 &38  &1.8&190  &0.081 & 3.7          &162&162&80.0&80.8&81.9& 3.7         &11                 \\
0.375 &96 &0.969&75 &33  &2.2&690  &0.213 & 1.4          &163&165&83.4&81.0&80.9& 1.4         &5.9                \\
0.500 &192&0.969&60 &27  &2.9&1700 &0.631 & 0.5          &164&170&94.7&96.2&96.8& 0.5         &$3.8^{\star}$      \\  
0.625 &96 &0.968&57 &25  &4.6&4500 &1.011 & 0.2          &166&164&80.3&83.1&83.3& 0.2         &$2.7^{\star}$     \\
0.750 &96 &0.966&49 &22  &8.4&11000&2.151 & 0.08         &171&162&77.8&86.1&85.1& 0.08        &$2.0^{\star}$      \\
0.875 &96 &0.963&48 &21  &18 &20000&4.651 & 0.03         &183&157&67.4&93.6&89.7& 0.03        &$1.6^{\star}$     \\
\hline
\end{tabular}
\vspace*{0.5cm}
\caption[]{Various properties {\em vs.} monomer density $\rho$ for chains of length $N=1024$:
the number of chains per box $M$,
the root-mean-square bond length $l$,
the root-mean-square chain end-to-end distance $\Rend(N)$ and the gyration radius $\Rgyr(N)$ discussed in sect.~\ref{res_RNRs},
the total non-bonded interaction energy $\enb$ per monomer and 
the interchain monomer interaction energy $\einter$ (Sect.~\ref{res_energy}),
the total pressure $P$ (Sect.~\ref{res_pressure}), 
the dimensionless compressibility $\gTN \equiv \kBT \rho / K$  (Sect.~\ref{res_compress}),
the hypervirial $\Xvirial$ and the excess pressure fluctuation $\Xfluctu$ (Sect.~\ref{res_moduli}),
and the contributions $\laB \equiv \muB$, $\laF$ and $\muF$ to the Lam\'e coefficients $\lambda$ and $\mu$ 
discussed in the Appendix. 
The last two columns refer to the dimensionless compressibility $\gT$ and the blob size $\xi$ for asymptotically long chains.
We assume that eq.~(\ref{eq_blobsize}) holds for all densities. Extrapolated values are indicated by stars ($\star$).
%
%
The thermodynamic properties are made dimensionless rescaling them with $\beta\equiv 1/\kBT$ and the number density $\rho$.
Note that the semidilute blobs become too small around and above $\rho \approx 0.5$ 
and that for $N=1024$ we have a crossover density $\rhostar \approx 0.85/N^{1/2} \approx 0.03$,
i.e. the semidilute regime ranges over about an order of magnitude in density.
\label{tab_rho}}
\end{table}

\end{document}